\documentclass[showpacs,preprint,nofootinbib,superscriptaddress,citeautoscript,eqsecnum,12pt]{article}
\usepackage{latexsym}
\usepackage{amsmath}
\usepackage{amsfonts}
\usepackage{amssymb}
\usepackage{amscd}
\usepackage{bbm}
\usepackage{fancybox}
\usepackage{cite}
\usepackage{amsmath,amsfonts,amsbsy}
\usepackage{pstricks,pst-node}
\usepackage[small,bf,hang]{caption2}
\usepackage{textcomp}
\usepackage{mathrsfs}
\usepackage{multirow}
\usepackage{array}
\usepackage[utf8]{inputenc}

\pdfoutput=1
\usepackage{graphicx}
\usepackage{color}
\usepackage{epsfig}
\usepackage{psfrag}
\usepackage{comment}
\usepackage{epstopdf}
\usepackage{hyperref}

\usepackage{relsize}
\usepackage{dcolumn}

\usepackage{float}

\psset{unit=1.3cm,linewidth=.5pt,radius=.2}  

\usepackage{multirow}                     
\usepackage{float}                          
\usepackage{lscape}                         
\usepackage{subfigure}          
\usepackage{bm}

\newcommand{\cL}{{\cal L}}
\newcommand{\cM}{{\cal M}}
\newcommand{\cN}{{\cal N}}

\newcommand{\cP}{{\cal P}}

\newcommand{\cW}{{\cal W}}

\newcommand{\bb}{\bar\beta}

\newcommand{\beq}{\begin{equation}}
\newcommand{\eeq}{\end{equation}}
\newcommand{\bi}{\begin{itemize}}
\newcommand{\ei}{\end{itemize}}
\newcommand{\bt}{\begin{tabular}}
\newcommand{\et}{\end{tabular}}
\newcommand{\bc}{\begin{center}}
\newcommand{\ec}{\end{center}}

\def\one{{\hbox{ 1\kern-.8mm l}}}
\newcommand{\Dslash}{\not{\hbox{\kern-4pt $D$}}}
\newcommand{\pdslash}{\not{\hbox{\kern-2pt $\partial$}}}

\newcommand{\be}{\begin{equation}}
\newcommand{\ee}{\end{equation}}
\newcommand{\bea}{\begin{eqnarray}}
\newcommand{\eea}{\end{eqnarray}}
\newcommand{\ba}{\begin{array}}
\newcommand{\ea}{\end{array}}

\def\bbox{{\,\lower0.9pt\vbox{\hrule \hbox{\vrule height 0.2 cm
\hskip 0.2 cm \vrule height 0.2 cm}\hrule}\,}}
\newcommand{\dsl}{\pa \kern-0.5em /}

\newcolumntype{C}[1]{>{\centering}m{#1}}

\newcommand{\tr}{\mathrm{tr}}

\newcommand{\vp}{\varphi}

\newcommand{\ve}{\varepsilon}

\font\mybb=msbm10 at 12pt
\def\bb#1{\hbox{\mybb#1}}

\def\bR {\bb{R}}

\numberwithin{equation}{section}
\usepackage[normalem]{ulem}

\begin{document}

\begin{titlepage}
	\begin{center}
		
		
		\vskip 1.5cm

		

		\noindent

		{\Large \bf Asymptotic dynamics of AdS$_3$ gravity with two asymptotic regions}\\

		\vskip 1cm

		{\bf Marc Henneaux$^{a,b}$}, {\bf Wout Merbis$^a$} and {\bf Arash Ranjbar$^a$}

		\vskip 25pt
		
		$^{a}${\em  Physique Th\'eorique et Math\'ematique, Universit\'e Libre de Bruxelles and International Solvay Institutes,  Campus Plaine - CP 231, B-1050 Bruxelles, Belgium \vskip 5pt }
		$^{b}${\em  Coll\`ege de France, 11 place Marcelin Berthelot, 75005 Paris, France \vskip 5pt }
		{email: {\tt \{henneaux, wmerbis, aranjbar\}@ulb.ac.be}} \\
		\vskip 10pt

	\end{center}

	\vskip 1.5cm
	
	\begin{center} {\bf ABSTRACT}\\[3ex]
	\end{center}
	
\noindent
The asymptotic dynamics of AdS$_3$ gravity with two asymptotically
anti-de Sitter regions is investigated, paying due attention to the zero
modes, i.e., holonomies along non-contractible circles and their
canonically conjugates.   This situation covers the eternal black hole
solution. We derive how the holonomies around the non-contractible
circles couple the fields on the two different boundaries and show that
their canonically conjugate variables, needed for a consistent dynamical
description of the holonomies, can be related to Wilson lines joining
the boundaries. The action reduces to the sum of four free chiral
actions, one for each boundary and each chirality, with additional non-trivial 
couplings to the zero modes which are explicitly written.   While
the Gauss decomposition of the $SL(2,\mathbb{R})$  group elements is
useful in order to treat hyperbolic holonomies, the Iwasawa
decomposition turns out to be more convenient in order to deal with
elliptic and parabolic holonomies. The connection with the geometric
action is also made explicit. Although our paper deals with the specific
example of two asymptotically anti-de Sitter regions, most of our global
considerations on holonomies and radial Wilson lines qualitatively
apply whenever there are multiple boundaries, independently of the form
that the boundary conditions explicitly take there. 
	
\end{titlepage}

\tableofcontents

\section{Introduction}
\label{sec:Intro}

Gravity in three-dimensional asymptotically anti-de Sitter (AdS)
spacetimes and two dimensional conformal field theories (CFT) have a
long common history, predating even the AdS/CFT correspondence
\cite{Deser:1984dr,Jackiw:1985je,Brown:1986nw,Achucarro:1987vz,Witten:1988hc,Verlinde:1989ua,Carlip:1991zm,Carlip3,Coussaert:1995zp}.
Rather than being derived from a top-down approach as in string theory,
the early relationship between gravity in 2+1 dimensions and CFTs was
based on semi-classical gravitational arguments, such as asymptotic
symmetry analysis. The relationship between $AdS_3$ gravity and
Liouville theory was made explicit in \cite{Coussaert:1995zp} and
generalized to supersymmetry in \cite{Henneaux:1999ib}. These works are
based on a Hamiltonian reduction of the Chern-Simons theory for
$SL(2,\mathbb{R}) \times SL(2,\mathbb{R})$ and supersymmetric
generalizations thereof, under the $AdS_3$ boundary conditions devised
in \cite{Brown:1986nw}.

From the point of view of Chern-Simons theory on manifolds with the
topology  $\mathcal{M}_2 \times \mathbb{R}$ (with $\mathcal{M}_2$ the
spatial sections and $\mathbb{R}$ the time), the first step of the
reduction utilizes the relation between Chern-Simons theory and two
chiral $SL(2,\mathbb{R})$ Wess-Zumino-Witten (WZW) models, well known
from \cite{Witten:1989hf,Moore:1989yh,Elitzur:1989nr}. Imposing the
standard $AdS_3$ boundary conditions was then shown in
\cite{Coussaert:1995zp} to be equivalent to the implementation of a
Drinfeld-Sokolov (DS) reduction \cite{Drinfeld:1984qv} of the WZW models.

The resulting action contains, in addition to local fields on the
boundaries,  global ``zero modes'' that are shared by the boundaries. 
These zero modes are the holonomies around non-contractible circles and
their canonically conjugates.  While the form of the action for the
boundary fields depends only on the boundary conditions at the
corresponding boundaries, with the property that the action on a given
boundary does not depend on the local fields on the other boundaries, 
the zero modes do couple the various boundaries. The exact form of the
coupling of the boundaries due to zero modes depend on the number of
boundaries, the topology of space and the boundary conditions at each
boundary.  It cannot therefore be addressed completely if one focuses
only on a single boundary without the full information.  This is the
reason why zero modes were not treated in the articles
\cite{Coussaert:1995zp,Henneaux:1999ib}, the purpose of which was to
understand the universal form of the symmetry algebra on a boundary with
AdS-type boundary conditions,  independently of the number and
properties of the other boundaries or of the topology.  Paper
\cite{Coussaert:1995zp} did not treat them at all (as explicitly
mentioned there), while paper \cite{Henneaux:1999ib} kept them up to the
point where the extra information is needed (see the appendix of that
reference).

\begin{figure}
	\hspace{-1cm}
	\includegraphics[width = 1.18\textwidth]{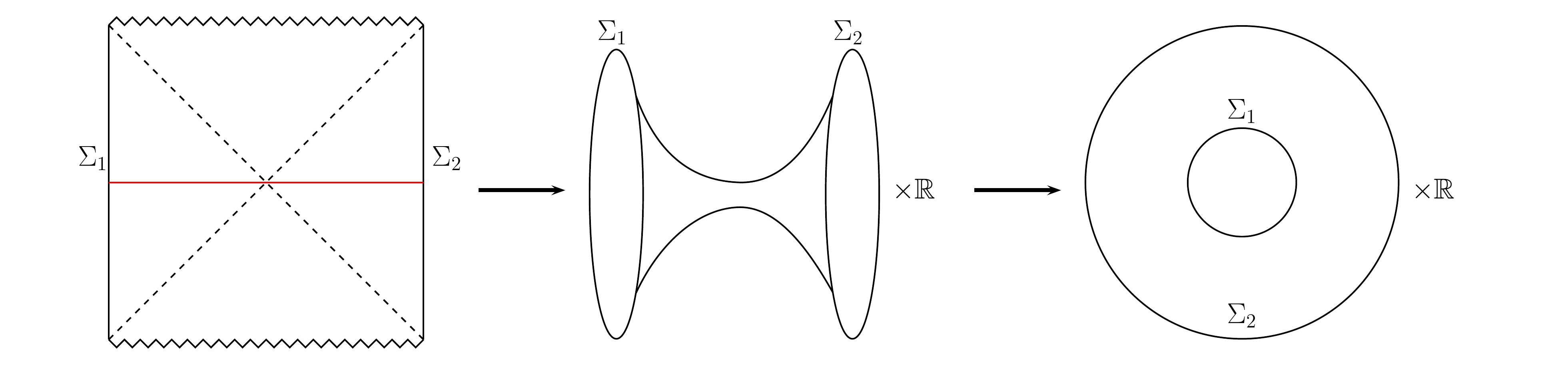}
	\caption{\label{fig1} A fixed time slice of the eternal black hole solution is an infinite cylinder, which is topologically equivalent to the annulus.   The two asymptotically AdS$_3$ boundaries are mapped on the two boundaries of the annulus. }
\end{figure}

The holonomy is trivial and non-dynamical only for manifolds of topology
``disk $\times$ time''.  This situation eliminates for instance the black
hole solutions \cite{Banados:1992wn,Banados:1992gq}.  It is therefore
necessary to go beyond that simple case.
In order to allow non-trivial holonomies, one must consider other
topologies, such as the annulus $\times$ time. It is actually exactly this
topology which is relevant for describing the three-dimensional black
holes. The eternal three-dimensional black holes have two asymptotic
regions where standard $AdS_3$ boundary conditions hold and can support
non-trivial holonomies along the non-contractible closed paths. Since
three-dimensional gravity is topological, we are free to deform the
manifold to our liking, as long as we do not change topology or boundary
conditions. Any fixed time-slice of the eternal BTZ black hole can then
be deformed as shown in figure \ref{fig1} to look like an annulus with boundaries at finite values of
the ``radial'' coordinate, with one asymptotic boundary as the inner
boundary and another asymptotic boundary as the outer boundary.

The three-dimensional black hole provides therefore a specific system
where the zero modes can be handled precisely.  In this work we consider
this setup and completely carry the reduction of the theory taking
special care of the dynamics of the holonomy.  One finds that AdS$_3$
gravity can then be reduced to a theory on both boundaries with a
specific coupling to the global zero modes which we derive.

An interesting feature of the analysis is that the holonomy is dynamical
and possesses a conjugate momentum.  This conjugate momentum can be
related to the radial Wilson lines connecting the boundaries.\footnote{The radial Wilson line has been discussed previously in the context of the eternal BTZ black hole in \cite{Castro:2016ehj} as computing probe operators in a thermofield double state.} The phase
space of the system is therefore not just two copies of the phase spaces
of the boundary theories, but there is in addition the global dynamical
zero modes described by the holonomy and the radial Wilson lines.  In
the quantum theory, the Hilbert space does not factorize into the mere
tensor product of the boundary Hilbert spaces, but involves also the
zero modes. The non-trivial link  between the boundary Hilbert spaces
does not need to be implemented by hand since it follows from the
action, which is {\it not} given by mere multiple copies of one boundary
action, but possesses an extra piece coupling the dynamical global zero
modes to the boundary fields.

It turns out that the resulting reduced boundary theory can be viewed as
a dynamical theory of the boundary Virasoro charges, which, together
with the zero modes, completely capture the physics of the system. 
These  charges transform in the coadjoint representation of the Virasoro
group and the boundary dynamics takes place on the coadjoint orbits.

This leads to two different descriptions of the boundary dynamics. One
is in terms of chiral bosons, which can be viewed as providing
Darboux-like coordinates on the coadjoint orbits.  Since there are two
$SL(2, \mathbb{R})$ factors and two boundaries, one ends up with four
chiral bosons theories, one per $SL(2, \mathbb{R})$ factor and
boundary.  We show that the chiral bosons corresponding to the same
$SL(2, \mathbb{R})$  at two different boundaries are linked by the fact
that the conjugate momentum to the holonomy (which is the same at the
two boundaries) is the difference between the zero modes of the chiral
fields at the two boundaries.

One can  combine the fields corresponding to the two different $SL(2,
\mathbb{R})$ factors at the same boundary into non-chiral Liouville
fields, as in \cite{Coussaert:1995zp,Henneaux:1999ib}, but the couplings
to the holonomies take then a more intricate form. As we shall explain, 
it turns out to be more natural and advantageous, however, to combine 
instead the fields corresponding to the same subgroup $SL(2,\mathbb{R})$ 
at the two different boundaries of the annulus (when the boundary conditions 
there have opposite chiralities, see below), because the fields share the same zero modes.

The other description of the reduced action is obtained by parametrizing
the boundary fields in terms of Virasoro group elements, leading to the
geometric action on the coadjoint orbits of the Virasoro group.  That
the Drinfeld-Sokolov reduction of the $SL(2,\mathbb{R})$ WZW model gives
the geometric action on the coadjoint orbit of the Virasoro group was
shown long ago in \cite{Alekseev:1988ce,Alekseev:1990mp}, and discussed
in the $AdS_3$ context in \cite{NavarroSalas:1999sr}. This latter
approach was recently revived  in
\cite{Barnich:2017jgw,Cotler:2018zff}.  With two boundaries, one gets
two copies of the geometric action per $SL(2, \mathbb{R})$, coupled through the fact that it is
the same holonomy (which characterizes the coadjoint orbits) that
appears in both of them.  The holonomy is one of the dynamical variables
and changes under the canonical transformation generated by its
conjugate momentum. In the quantum theory, the path integral would involve a sum over the orbits.

Our paper is organized as follows:
As a warm-up exercise, we review first in Section \ref{sec:U1_Chern_simons} the simplest case
of a $U(1)$ Chern-Simons theory, which illustrates the main point. This
is in fact standard material developed in
\cite{Witten:1989hf,Moore:1989yh,Elitzur:1989nr}, where a more profound
analysis going much beyond the considerations below can be found.  We
explicitly construct the conjugate variable to the non-trivial holonomy
in the annulus case and show how it is  simply related to the radial
Wilson lines connecting the two boundaries. The connection with the non-chiral 
free boson on the cylinder is then established. We also discuss the
extension to the non-abelian case, and in particular how the abelian
condition that the holonomy is the same at both boundaries becomes a
matching condition expressing that the boundary holonomies are in the
same conjugacy class.  We further recall that the radial Wilson lines
connecting the boundaries have now more complicated Poisson brackets
with the boundary fields and the holonomy, but remain crucially related
to the conjugate momenta to the holonomy in the sense that radial Wilson
lines and holonomies around the non-trivial circles have non-vanishing
Poisson brackets. A complete description of the system is given by the
boundary Kac-Moody algebras (constrained by the matching condition that
they should define holonomies in the same conjugacy class) and one
radial Wilson line. 

We then turn to gravity in Section \ref{sec:gravity}, where in addition to the 
Chern-Simons reduction to a WZW model at the boundary, the asymptotic 
conditions impose the Drinfeld-Sokolov Hamiltonian reduction 
\cite{Coussaert:1995zp,Henneaux:1999ib}.   As shown in 
\cite{Witten:1989hf,Moore:1989yh,Elitzur:1989nr}, the reduction of the 
Chern-Simons theory to a boundary chiral WZW theory involves a gauge 
redundancy.  The DS reduction conditions, which are expressed in terms 
of gauge invariant currents, preserve this gauge invariance.  Thus, the 
reduction of $AdS_3$ gravity on the annulus, leading to two boundary 
theories coupled through the zero modes (holonomy and radial Wilson 
line), has that gauge symmetry.  As explained in the appendix of
\cite{Henneaux:1999ib}, the gauge symmetry can be partly 
fixed by choosing a particular form of the holonomy.  The residual gauge 
symmetry will then be given by the gauge transformations that preserve 
the form of the holonomy.

Now, the  holonomy  in $SL(2, \mathbb{R})$ can be of three different 
types: hyperbolic, elliptic or parabolic.  The (partly) gauged fixed 
form of the holonomy will be different in each case. In the hyperbolic 
case, considered in \cite{Henneaux:1999ib}, the holonomy can be assumed 
to be diagonal (exponential of a Cartan subalgebra element).  This case, 
which covers the BTZ black hole, is treated in full in Subsection \ref{sec:hyperbolic}, 
where the Gauss decomposition of the $SL(2, \mathbb{R})$ group elements 
is found particularly convenient.  We compute in particular the radial 
Wilson line for the non-rotating black hole and show that its non-trivial 
value can be viewed as an obstruction to going to the gauge $A_r 
=0$ while preserving the DS reduction conditions at the two boundaries 
(which are ``twisted'' with respect to one another, one taking the 
highest weight form and the other the lowest weight one).

For the other types of holonomies, another classical decomposition of 
matrices is found to be convenient, namely the Iwasawa decomposition.  
Elliptic holonomies are dealt with in Subsection \ref{sec:elliptic}, where we emphasize 
in particular the enlargement of the residual gauge symmetry when the 
holonomy is in the center of $SL(2, \mathbb{R})$ (for which the gauge is 
in fact not fixed at all).  The parabolic holonomy case is considered in 
Subsection \ref{sec:parabolic}.

Section \ref{sec:Conclusions} is devoted to conclusions and comments. 
We mention there the extension to supergravity and higher-spin gauge 
theories, where the fact that the asymptotic symmetry algebras are 
non-linear forces one to formulate the geometric actions in terms of 
symplectic leaves,  which generalize the concept of coadjoint orbits 
when the Poisson manifold of the boundary currents is endowed with a 
nonlinear Poisson bracket structure. The paper closes with four appendices 
of a more technical nature.

\section{$U(1)$ Chern-Simons theory}\label{sec:U1_Chern_simons}

We start with the Chern-Simons action for a single abelian field $A_\mu$ given by
\be
S_{[CS]} = \frac{k}{2 \pi}  \int dt dr d \varphi   \left(  A_\varphi \partial_t A_r + A_t F_{r \varphi} \right) \, ,
\ee
up to boundary terms which are discussed below. The topology is $\mathbb{R} \times M$ where $M$ is a two-dimensional manifold with coordinates $(r, \varphi)$.  

The kinetic term in the action shows that $A_\varphi$ and $A_r$ are canonically conjugate in the Poisson bracket
\be
[A_r (r,\varphi), A_\varphi(r',\varphi')] = \frac{2 \pi}{k} \delta(r-r') \delta(\varphi - \varphi')\,,
\ee
(in general coordinates, this is $[A_i(\mathbf{x}), A_j(\mathbf{y})]= \frac{2 \pi}{k} \varepsilon_{ij} \delta^{(2)}(\mathbf{x} - \mathbf{y})$, which is coordinate invariant).

The $A_t$-equation of motion implies the constraints $F_{r \varphi} = 0$, from which one gets
\be
A_r = \partial_r \Lambda \,,
\ee
and
\be
A_\varphi = \partial_\varphi \Lambda + k  \,,
\ee
with $\Lambda = \Lambda(r, \varphi)$ and $k  = k (\varphi)$ single-valued functions.  These are also functions of time, but we do not write systematically the $t$-dependence.

\subsection{Proper and improper gauge symmetries}

The constraint generates gauge transformations, in the sense that
\be
Q[\epsilon] = - \frac{k}{2 \pi} \int d^2x \,\epsilon\, F_{r \varphi} + B_{\partial M} \,,
\ee
is such that
\be
\delta A_r = [A_r, Q[\epsilon]] = \partial_r \epsilon, \qquad \delta A_\varphi = [A_\varphi, Q[\epsilon]] = \partial_\varphi \epsilon.
\ee
Here, $B_{\partial M}$ is a boundary term that must be added to the bulk term in order for the transformation $\delta A_r = \partial_r \epsilon $, $\delta A_\varphi =  \partial_\varphi \epsilon$, which is canonical (i.e., which leaves the symplectic $2$-form invariant, $d_V (i_X \sigma) = 0$) to indeed be generated by $Q[\epsilon]$, (i.e., $i_X \sigma = - d_V Q[\epsilon]$ exactly, and not just up to surface terms \cite{Regge:1974zd,Henneaux:2018gfi})\footnote{We follow the notations of \cite{Henneaux:2018gfi}: $d_V$ is the exterior derivative in field space; $X$ is the vector field in field space defined by the infinitesimal transformations; $i_X \sigma$ is the inner contraction of the symplectic form $\sigma$ by $X$.  Note that the Lie derivative $\mathcal{L}_X \sigma$ of $\sigma$ reduces to $\mathcal{L}_X \sigma = (d_V i_X + i_X d_V) \sigma = d_V i_X \sigma$ because $\sigma$ is closed.}.  It reads
\be
B_{\partial M} = \frac{k}{2 \pi} \oint_{\partial M} d \lambda \, \epsilon(\lambda)\vert_{\partial M}  \, A_\lambda(\lambda)\vert_{\partial M}\,,
\ee
where $\lambda$ is a coordinate on the boundary. We have assumed that $\epsilon$ is field-independent.

So, in the case of a disk with boundary at $r = r_1$, the surface term reads
\be
B_{\partial M} = \frac{k}{2 \pi} \oint d \varphi \, \epsilon(r=r_1, \varphi)  \, A_\varphi(r=r_1, \varphi)\,,
\ee
while for an annulus $r_1 \leq r \leq r_2$, it becomes
\be
B_{\partial M} = \frac{k}{2 \pi} \oint d \varphi \, \epsilon(r=r_2, \varphi)  \, A_\varphi(r=r_2, \varphi) - \frac{k}{2 \pi} \oint d \varphi \, \epsilon(r=r_1, \varphi)  \, A_\varphi(r=r_1, \varphi).
\ee

Now, there are two types of ``gauge transformations'' \cite{Benguria:1977in}: ``proper ones'' that correspond to mere redundancies, and ``improper ones'' that do change the physical states of the system.  What distinguishes the two are the values of the generators, which reduce on-shell to the boundary terms.  Proper gauge transformations have generators that vanish (on-shell) for all configurations under consideration.  By contrast, the generators of improper gauge transformations need not vanish (even when the constraints hold).  This can clearly happen only when the gauge parameter $\epsilon$ does not vanish at the boundary.

A direct computation shows that the algebra of the charges $Q[\epsilon]$ is given by
\be
[Q[\epsilon], Q[\eta]] = \frac{k}{2 \pi} \oint_{\partial M}  d \lambda \epsilon(\lambda)\vert_{\partial M}  \frac{d}{d \lambda} \eta(\lambda)\vert_{\partial M} \,.  \label{eq:QQ}
\ee
The easiest way to check this relation is to observe that $[Q[\epsilon], Q[\eta]] = \delta_\eta Q[\epsilon]$, and use the known gauge  transformation rule of the vector potential at the boundary, which is the only quantity that transforms in  $Q[\epsilon]$ under $Q[\eta]$.  

It follows from this relation that $ [Q[\epsilon], Q[\eta]] =0$ whenever $\eta$ vanishes at the boundary.  This means that $Q[\epsilon]$ is invariant under proper gauge transformations, i.e. an observable of the theory, which is  non-trivial when $\epsilon$ does not vanish at the boundary.  By expanding the boundary gauge parameter $\epsilon(\lambda)$ in terms of a basis of functions on $\partial M$, one gets an infinite number of observables.

Two questions arise then: (i) Are these observables unconstrained? (ii) Are they complete?  The answers to both questions depend on the topology of space.  This issue will be addressed in the next sections.

It is sometimes convenient to fix the gauge.  A good gauge choice eliminates the redundancy of proper gauge transformations without factoring out the improper gauge transformations.  After the gauge is fixed, the constraints can be used as strong equations and the bracket to be used is the Dirac bracket.  The relation (\ref{eq:QQ}) is equivalent to
\be
[A_\lambda(\lambda), A_\lambda(\lambda')] = \frac{2 \pi}{k} \partial_{\lambda} \delta(\lambda - \lambda')\,.
\ee
The bracket is here the Dirac bracket.  This is the familiar $u(1)$ Kac-Moody algebra, with the tangential components of the vector potential being the Kac-Moody currents.

\subsection{Zero modes for disk topology (a single boundary)}

\subsubsection{Action}
We now turn to specific examples to illustrate the key features of the above discussion.  We consider first the case of a disk.
We assume that the disk is centered at the origin and has radius $r_1$.  There is a single boundary at $r=r_1$.  The above boundary parameter is taken to be equal to $\varphi$.

In this case, $\oint d \varphi A_\varphi = 0$ for all values of $r$, since the circles $r =$ const. are contractible.  This implies $\oint d \varphi k = 0$, from which one derives $ k = \partial_\varphi \lambda$ with $\lambda$ single-valued.  We can thus write
\be
A_r = \partial_r \mu, \; \; \; 
A_\varphi = \partial_\varphi \mu  \,,   \label{eq:SolForA01}
\ee
with $\mu = \Lambda + \lambda$.

This parametrization of $A_i$ has some redundancy since 
$ \mu \rightarrow \mu + \epsilon$
 with $\epsilon = \epsilon(t)$ (writing explicitly the time dependence) leaves $A_i$ unchanged.  This is a gauge symmetry.
 
As discussed in \cite{Coussaert:1995zp}, the boundary condition relevant to $AdS_3$ gravity with $AdS$ boundary conditions is\footnote{Note a typo in (2) of  \cite{Coussaert:1995zp}, where the last term should be $+2 A_0  F_{r \varphi}$, instead of $- A_0  F_{r \varphi}$. The subsequent discussion of the boundary terms has the correct factors.   Note also that the boundary condition $A_-=0$ should be viewed as a condition that expresses $A_0$ in terms of $A_\varphi$, which is not restricted by it. The choice $A_0 = A_\varphi$ is an (improper) gauge choice, which selects a definite evolution of the initial data.  The time evolution in Chern-Simons theory is indeed a gauge transformation with gauge parameter $A_0$. The choice $A_- = 0$ leads to a simple evolution (as does $A_+ = 0$). With $A_0 = A_\varphi$, the gauge parameter in front of the constraints depends on $A_\varphi$, which explains why there is a factor of $1/2$ in front of $(A_\varphi)^2$ in $H$, $A_\varphi \delta A_\varphi = \delta (\frac12 A_\varphi^2)$, as shown in the text.   One could consider more general $A_0$'s, which can generate an arbitrary boundary symmetry, leading to different generators $H$ of the dynamical evolution at the boundary. }   $A_- = 0$ with $A_- = A_t - A_\varphi$, i.e., $A_t = A_\varphi$.   Under this condition, the variation of the action picks up the boundary term at $r = r_1$
\be
 \frac{k}{2 \pi} \int dt \oint d \varphi A_t \delta A_\varphi \biggr\vert_{r = r_1} = \frac{k}{4 \pi} \int dt \oint d \varphi\, \delta(A_\varphi^2)\biggr\vert_{r = r_1} \,,
\ee
 which must be compensated by adding the boundary term at $r=r_1$
 \be 
 - \frac{k}{4 \pi}  \int dt H \,,
 \ee
 with 
 \be
 H = \oint d \varphi \left( A_\varphi \right)^2 \biggr\vert_{r = r_1}\,.
 \ee
 
 Plugging the form of $A_i$ in the action (with the surface term included at the spatial boundary) and dropping a surface term at the time boundaries, one gets the free chiral boson action \cite{Floreanini:1987as}
 \be
S[\Phi(t, \varphi)] = \frac{k}{4 \pi}  \int dt \left[ \oint d \varphi   \left(  \partial_\varphi \Phi \partial_t \Phi  \right) - H \right], \qquad H = \oint d \varphi \left( \partial_\varphi \Phi \right)^2   \,, \label{eq:ActionDisk}
\ee 
where $\Phi$ is the value of $\mu$ at the boundary $r=r_1$, $\Phi(t, \varphi) \equiv \mu (t, r_1, \varphi)$.  

In agreement with the redundancy mentioned above, the action (\ref{eq:ActionDisk}) is  found to be invariant under the transformation 
\be
\Phi \rightarrow \Phi + \epsilon(t) \,.  \label{eq:GaugeSymmDisk}
\ee
This gauge symmetry shows that the zero mode of $\Phi$ is pure gauge and can be set equal to any value.  

The global symmetry on the boundary is generated by the Kac-Moody currents $ j \equiv \frac{k}{2\pi} A_\varphi$ which are equal to $A_\varphi = \partial_\varphi \Phi $ on the boundary, and reads
\be
\Phi \rightarrow \Phi +  \nu(\varphi) \,, \label{eq:RigidSymmDisk}
\ee
where $\nu(\varphi)$ is an arbitrary time-independent function of $\varphi$.   The zero mode of this transformation coincides with the gauge transformation (\ref{eq:GaugeSymmDisk}) and can depend also on time. Quotienting out this gauge symmetry, one sees that the actual global symmetry is thus $\widehat{LG}/G$ with  $G = U(1)$ \cite{Witten:1989hf,Moore:1989yh,Elitzur:1989nr}.  Note that the zero mode of the Kac-Moody currents, which is the holonomy,  is identically zero when expressed in terms of  $\Phi$.  

\vspace{0.3cm}

\noindent
{\bf Comments}
\begin{itemize}

\item The momentum $\pi_0$ conjugate to the zero mode is zero. The constraint $\pi_0 \approx 0$ is first class and generates the gauge symmetry.

\item In fact, since $U(1)$ is not simply connected, there are additional sectors  \cite{Moore:1989yh} but these will not be discussed here.

\end{itemize}

\subsubsection{Do the Kac-Moody currents form a complete set of observables?}

The Kac-Moody currents $j(\varphi) \equiv \frac{k}{2\pi} A_\varphi (\varphi)$  fulfill the bracket relations
\be
[j(\varphi), j(\varphi')] = \frac{k}{2 \pi}\, \partial_{\varphi}\delta(\varphi - \varphi')\, .   \label{eq:KM1}
\ee
They are constrained by the condition that their zero mode (the holonomy) be zero, a condition which is compatible with the algebra since $[\oint d \varphi \, j(\varphi), j(\varphi')] = 0$.

Although the bracket has been technically derived as a Dirac bracket, we shall often refer to it as the ``Poisson bracket'',  since it will be the fundamental bracket between the observables that will be the starting point for the geometric considerations. No confusion should arise.

Do the (constrained) currents form a complete set of observables? By this we mean: if we prescribe $j(\varphi)$ on the boundary at a given ``initial time'', are the initial data $(A_r(r, \varphi), A_\varphi(r,\varphi))$ completely specified up to proper gauge transformations?  This amounts to determining the general solution of $F_{r\varphi}=0$ with the prescribed boundary conditions.

We have seen that the general solution of the constraint is given by (\ref{eq:SolForA01}), i.e., $A_r = \partial_r \mu$ and $A_\varphi = \partial_\varphi \mu$ for some function $\mu(r, \varphi)$.  If  $A_\varphi (r_1, \varphi) $ (without zero mode) is given, the only allowed gauge transformations are
\be
\mu \rightarrow \mu + \Lambda(r,\varphi)\,,
\ee
where the gauge parameter is constrained to reduce to a constant at the boundary,
\be
\Lambda(r= r_1, \varphi) = C \,,
\ee
in order to match the given $A_\varphi(r=r_1, \varphi)$.
But this is a proper gauge transformation since the corresponding charge vanishes,
 \be
 \frac{k}{2 \pi} \oint d \varphi \, C  \, A_\varphi(r=r_1, \varphi) =  \frac{k}{2 \pi} C \oint d \varphi \, A_\varphi(r=r_1, \varphi) = 0\,.
 \ee
The solution is thus completely determined by the boundary value of $A_\varphi$ up to a proper gauge transformation.  The conclusion is therefore that the Kac-Moody currents at the boundary form a complete set of observables when the spatial section have the disk topology.  

To summarize: the dynamics is completely captured by the dynamics of the Kac-Moody currents $j(\varphi) \equiv \frac{k}{2\pi} A_\varphi(\varphi)$, subject to the constraint $\oint j(\varphi) =0$ and fulfilling the bracket (\ref{eq:KM1}).  The Hamiltonian $H$ is $(\frac{2\pi}{k})^2 \oint j^2(\varphi)$.  One can equivalently parametrize the current in terms of the chiral boson $\Phi(\varphi)$, which is unconstrained but has some gauge redundancy (zero mode).

\subsubsection{Connection with geometric action}

This is precisely the setting of the geometric formulation of the dynamics in terms of coadjoint orbits of the Kac-Moody group.

The currents (``charges''), which provide a complete physical description of the system as we have just seen, parametrize the vector space dual to the Lie algebra and transform in the coadjoint representation.  The transformation  is generated by the currents themselves acting through the canonical Poisson bracket (\ref{eq:KM1}) associated with the Lie algebra structure.  It preserves therefore the non-degenerate symplectic structure induced on the coadjoint orbits \cite{Kirillov1,Kirillov2,Kostant,Souriau,Khesin}.

In our case, there is only one relevant orbit, namely, the  orbit with zero holonomy (two connections $A_\varphi$ with same holonomy can be mapped on one another by a Kac-Moody transformation and the stability subgroup is just $U(1)$ as the formulas (\ref{eq:GaugeSymmDisk}), (\ref{eq:RigidSymmDisk})  show).  

The parametrization of the orbits in terms of $\Phi$ is adapted to the group action since the transformations are then just shifts of $\Phi$. The symplectic structure can be read off from the action and is $\oint d \varphi\, d(\partial_\varphi \Phi) \wedge d \Phi$.  It is non-degenerate when the zero mode of $\Phi$ is quotiented out.  The kinetic term is the so-called ``geometric action''.   

\subsection{Zero modes for annulus topology (two boundaries)}

\subsubsection{Action}

If the spatial manifold is an annulus with boundaries, say, at $r = r_1$ and $r = r_2$ ($r_1 < r_2$), the holonomy $\oint d \varphi A_\varphi$ need not vanish.  One  thus has $k = \partial_\varphi \lambda + k_0$, where $k_0$ does not depend on $\varphi$.  This yields
\be
A_r = \partial_r \mu, \; \; \; 
A_\varphi = \partial_\varphi \mu + k_0 \,, \label{eq:holonomyU(1)}
\ee
with $\mu = \Lambda + \lambda$.  There is again the redundancy $ \mu \rightarrow \mu + \epsilon$
 with $\epsilon = \epsilon(t)$.
 
 The boundary term picked up at the spatial boundaries reads now
 \be
 \frac{k}{2 \pi} \int dt \oint d \varphi A_t \delta A_\varphi \biggr\vert_{r = r_2} - \frac{k}{2 \pi} \int dt \oint d \varphi A_t \delta A_\varphi \biggr\vert_{r = r_1}\,.
\ee
In order to have a positive Hamiltonian, we should impose the condition $A_- = 0$ at the outer boundary $r = r_2$ as above, but $A_+ = 0$ ($\Leftrightarrow A_t = - A_\varphi$) at the inner boundary $r=r_1$, so that the boundary term becomes
\be
 \frac{k}{4 \pi} \int dt \oint d \varphi\, \delta(A_\varphi^2)\biggr\vert_{r = r_2} +  \frac{k}{4 \pi} \int dt \oint d \varphi\, \delta (A_\varphi^2)\biggr\vert_{r = r_1} \,,
\ee
to be canceled by adding to the action 
 \be 
 - \frac{k}{4 \pi}  \int dt H \,,
 \ee
 with 
 \be
 H = \oint d \varphi \left( A_\varphi \right)^2 \biggr\vert_{r = r_2} + \oint d \varphi \left( A_\varphi \right)^2 \biggr\vert_{r = r_1},  \quad (A_-(r = r_2) = 0, \quad A_+(r = r_1) = 0)\,.
 \ee
 The boundary condition $A_- = 0$ at the boundary $r = r_1$ is also worth being considered even though the corresponding Hamiltonian is not bounded from below. One then gets 
 \be
 H = \oint d \varphi \left( A_\varphi \right)^2 \biggr\vert_{r = r_2} - \oint d \varphi \left( A_\varphi \right)^2 \biggr\vert_{r = r_1}, \qquad (A_-(r = r_2) = 0, \quad A_-(r = r_1) = 0)\,.
 \ee
 The two situations $A_t = \pm A_\vp$ can be thought of as differing by the orientation of time at the inner boundary ($A_t \rightarrow - A_t$).

Inserting the expression for $A_i$ into the action, one gets with the first boundary conditions
\begin{subequations}\label{chiralboson}
\begin{eqnarray}
&& S[\Phi(t, \varphi), \Psi(t, \varphi), k_0(t)] = S^{(2)} + S^{(1)} + S^{(0)}, \label{eq:ActionAnnulus} \\
&& S^{(2)} = \frac{k}{4 \pi}  \int dt \left[ \oint d \varphi   \left(  \partial_\varphi \Phi \partial_t \Phi  \right) - H_\Phi \right], \qquad H_\Phi = \int d \varphi \left( \partial_\varphi \Phi \right)^2 , \\
&& S^{(1)} = \frac{k}{4 \pi}  \int dt \left[ - \oint d \varphi   \left(  \partial_\varphi \Psi \partial_t \Psi  \right) - H_\Psi \right], \qquad H_\Psi = \int d \varphi \left( \partial_\varphi \Psi \right)^2 , \\
 && S^{(0)} = \frac{k}{2 \pi} \int dt \left[ \oint d \varphi\, k_0  (\partial_t \Phi - \partial_t \Psi )  - H_0 \right], \qquad H_0 = 2 \pi \left(k_0 \right)^2\,,
\end{eqnarray} 
\end{subequations}
where $\Phi$ and $\Psi$ are the fields at the boundaries, $\Phi(t,\varphi) = \mu(t, r_2, \varphi)$, $\Psi(t,\varphi) = \mu(t, r_1, \varphi)$. 
For the second boundary conditions, one ends up with the same expression but now $ H_\Psi = -\int d \varphi \left( \partial_\varphi \Psi \right)^2$ and $H_0 = 0$.

In both cases, the action has the gauge symmetry
\be
\Phi \rightarrow \Phi + \epsilon(t), \qquad \Psi \rightarrow \Psi + \epsilon(t), \qquad k_0 \rightarrow k_0 \,, \label{eq:GaugeSymmAnnulus}
\ee
coming again from the redundancy of the parametrization of $A_i$. The zero mode of $\Phi - \Psi$ is gauge invariant and conjugate to the holonomy $k_0$ (up to a constant), which is a dynamical variable.  The phase space contains therefore configurations of the connection with different holonomies, while the holonomy was fixed to be zero in the previous section and not a dynamical variable.

More precisely, if we denote by $\pi_0$ and $p_0$ the momenta conjugate to the zero modes of $\Phi$ and $\Psi$, we get from the action $\pi_0 = k k_0$ and $p_0 = - k k_0$.  The fact that it is the same holonomy $k_0$ that appears in both expressions implies the first class constraint $\pi_0 + p_0 \approx 0$, which is actually the generator of the gauge transformation (\ref{eq:GaugeSymmAnnulus}).  At the same time, the momentum $\Pi_0$ conjugate to the holonomy $k_0$ is explicitly equal to 
\be \label{U1action}
- \Pi_0 = \frac{k}{2 \pi} \oint d \varphi (\Phi - \Psi) \, ,
\ee 
since the relevant term in the kinetic term reads $-\int dt k_0 \dot{\Pi}_0$. There is only one zero mode, which is shared by the two boundaries due to the condition $\frac{d}{dr}\oint d \varphi A_\varphi = 0$ that follows from the zero curvature condition (the second zero mode is pure gauge).

The global symmetry is generated by the two sets of Kac-Moody currents $j(\varphi)$ at the outer boundary and $m(\varphi)$ at the inner boundary defined as
\be
j(\varphi) \equiv \frac{k}{2\pi} A_\varphi^{(2)}(\varphi), \qquad m(\varphi) \equiv \frac{k}{2\pi} A_\varphi^{(1)}(\varphi)\,.
\ee
The global symmetry is in fact $\frac{\widehat{LG} \times \widehat{LG}}{G}$, with one copy of $\widehat{LG}$ at each boundary.  The quotient by $G$ accounts for the fact that the two algebras share the same zero mode,  $\oint d \varphi j(\varphi) = \oint d \varphi m(\varphi) = k_0$. [If one changes the orientation of the inner boundary so that the normal points outwards at both boundaries, one finds $-k_0$ as holonomy at the inner boundary, so that the sum of the boundary holonomies is zero.]

It is easy to check that the equations of motion imply that the holonomy is time-independent,
\be
\dot{k}_0 = 0 \,.
\ee
Depending on the choice of boundary conditions at the inner boundary,  its conjugate momentum $\Pi_0$ either grows linearly with time (first choice, $H_0 = 2 \pi (k_0)^2$) or remains constant (second choice, $H_0 = 0$).

The two boundaries are coupled through the kinetic term in $S^{(0)}$, which involves the zero modes.  One may decouple them in the abelian case under study by choosing the gauge $\oint d \varphi \Psi = 0$.  The action on the outer boundary $r = r_2$ becomes 
 \be
S[\Phi(t, \varphi), k_0(t)] = \frac{k}{4 \pi}  \int dt \left[ \oint d \varphi   \left(  \partial_\varphi \Phi \partial_t \Phi  \right) + 2 k_0 \oint d \varphi \, \partial_t \Phi - H \right] ,  \label{eq:ActionAnnulus2}
\ee
with
\be
H = \int d \varphi \left( \partial_\varphi \Phi \right)^2  +  4 \pi \left(k_0 \right)^2\, .
\ee
It has no gauge symmetry and possesses global symmetry $\widehat{LG}$.  It differs from the action for the disk by the contribution of the zero modes.  The zero mode $k_0$ of the Kac-Moody current (holonomy) is now not constrained to vanish and is conjugate to the zero mode of $\Phi$, which is not anymore pure gauge.  

In this gauge, the action at the inner boundary is the same as the action for the disk and has symmetry $\widehat{LG}/G$, but the zero mode of the Kac-Moody current is constrained to be equal to $k_0$ (instead of $0$).

We can thus conclude that the description of the dynamics at the outer boundary depends, in what concerns the zero modes, ``on what is inside''.   As announced above, extra information on the presence of other boundaries and on the topology is necessary to control the zero modes.

 Note also that an equivalent way to treat the holonomy on the annulus is to insert a dynamical source in a space with the disk topology \cite{Elitzur:1989nr}.
 
\subsubsection{Connection with non-chiral boson}
 
The action $S[k_0, \Phi, \Psi]$ given by (\ref{eq:ActionAnnulus}) contains two chiral bosons of opposite chiralities, as well as a common zero mode that they share.  By making the change of variables 
 \be
 \phi = \Phi - \Psi\,, \qquad \Pi_\phi = \frac{k}{2 \pi} \left(\Phi' + \Psi' + 2 k_0\right)\, , \qquad a = \Phi_0 + \Psi_0 \,,
 \ee
 which is invertible, one can rewrite it as
 \be
 S[\phi, \Pi_\phi] = \int dt \left(\int d \vp  \Pi_\phi \dot{\phi} - H \right) \,, \label{HamActionNonChiral}
 \ee with
 \be
 H = \int d \vp  \left[\frac{4\pi}{k} \Pi_\phi^2 + \frac{k}{16 \pi} \phi'^2 \right]\,.
 \ee
 The variable $a$ drops out because it is pure gauge and so, one can forget about it in the action (\ref{HamActionNonChiral}), which has then no gauge invariance left.
 
 The action (\ref{HamActionNonChiral}) is just the Hamiltonian form of the action for a non-chiral boson. Eliminating the conjugate momenta through their own equations of motion leads to the standard action 
 \be
 S[\phi] = \frac{k}{16 \pi} \int dt d\vp \left[(\partial_t \phi)^2 - (\partial_\vp \phi)^2 \right] \,,
 \ee
 for a free boson on a cylinder.
 
We thus see that with the boundary condition $A_- = 0$ at one boundary and the boundary condition $A_+= 0$ at the other boundary, the resulting theory is just that of a non-chiral free boson.  Crucial in the reconstruction is the fact that the chiral bosons at the two boundaries have opposite chiralities and share the same zero mode.  With the boundary conditions $A_- = 0$ at the two boundaries, the two chiral bosons would have the same chiralities and the above construction could not be applied.

\subsubsection{Do the Kac-Moody currents form a complete set of observables?}

The Kac-Moody currents fulfill the bracket relations
\begin{subequations}
\begin{eqnarray}
&& [j(\varphi), j(\varphi')] = \frac{k}{2 \pi} \partial_\vp\delta(\varphi - \varphi')\,,   \label{eq:KM2a} \\
&& [m(\varphi), m(\varphi')] = \frac{k}{2 \pi} \partial_\vp\delta(\varphi - \varphi')\,,   \label{eq:KM2b} \\
&& [j(\varphi), m(\varphi')] = 0 \,,   \label{eq:KM2c}
\end{eqnarray}
\end{subequations}
which follow from the kinetic term in the action, and obey the constraint
\be
\oint d \varphi\, j(\varphi) = \oint d \varphi\, m(\varphi) = k_0\, .
\ee

In the case of the disk topology, the boundary Kac-Moody currents provided a complete description of the system.  Is the same true in the case of an annulus?  Does the knowledge of the two boundary Kac-Moody currents $(j(\varphi), m(\varphi))$ completely determine the gauge potential $A_i$ on the annulus up to a proper gauge transformation?

We know that the answer is negative, since the conjugate momentum $\Pi_0$ to the holonomy is gauge invariant and not determined by the boundary Kac-Moody currents (contrary to the holonomy).   To completely specify the classical state of the system up to irrelevant proper gauge transformations, one needs to give not only the two sets of Kac-Moody currents at the boundaries (subject to equal zero mode) but also $\Pi_0$.
 
 One way to understand this is to observe that $\Phi-\Psi$ can be written as
 \be
 \Phi - \Psi = \int_{r_1}^{r_2} dr A_r \,,
 \ee
a quantity that is manifestly gauge invariant under all proper gauge transformations; it is a Wilson line along a radial curve connecting the two boundaries.  The momentum conjugate to the holonomy can thus be expressed non-locally in terms of the radial component of the gauge potential,
\be
 \Pi_0 =  - \frac{k}{2 \pi} \oint d \varphi \left(\int_{r_1}^{r_2} dr A_r \right)\,.
\ee 
While the Fourier modes $(\Phi - \Psi)_n$ of $\Phi - \Psi$ with $n \not=0$ are completely determined by the boundary currents, this is not true for the zero mode $\Pi_0$, which involves independent bulk data through the integral of the vector potential along lines joining the two boundaries.  It is amusing to note that the radial component $A_r$ of the Maxwell field plays a similar role at infinity in the asymptotically flat context \cite{Henneaux:2018gfi}. Note that instead of integrating along a radial direction, one can take the integral along any curve joining the two boundaries.  This will amount to shifting the momentum $\Pi_0$ by a multiple of $k_0$, which is a canonical transformation.

The time derivative of $\Pi_0$ follows from the equation $F_{tr}= 0 \Leftrightarrow \partial_t A_r = \partial_r A_t$,
\be
\dot{\Pi}_0 = - \frac{k}{2 \pi} \oint d \vp \left(\int_{r_1}^{r_2}  \partial_r A_t \right) = -  \oint d \vp \left( j(\vp) \pm m(\vp) \right)\,,
\ee
depending on whether $A_t = \mp A_\vp$ at the lower boundary.  Using the holonomy matching condition between lower and upper boundaries we get $\dot{\Pi}_0 = - 2 k_0$, or $\dot{\Pi}_0 = 0$, in agreement with the Hamiltonian equation.

Finally, we observe that the brackets relation involving $\Pi_0$ read
\be
[\Pi_0, j(\varphi)] = - 1 , \qquad [\Pi_0, m(\varphi)] = - 1, \label{eq:Heisenberg}
\ee
implying in particular $[\Pi_0, j_n] = 0$ and  $[\Pi_0, m_n] = 0$ for $n \not=0$.

One can again describe the dynamics of the fields at one boundary in terms of coadjoint orbits of the Kac-Moody group  and geometric actions.  But there is an additional feature compared with the disk topology, namely, that the holonomy $k_0$, which is a constant of the motion, was previously fixed to be zero but can now take arbitrary values.  Furthermore,  there is a matching condition between the geometric dynamical description at one boundary and the geometric dynamical description at the other boundary, namely, that the holonomies are equal (or differ by the sign if one changes the orientation at the inner boundary). 

One can alternatively consider the coadjoint orbits of the full algebra (\ref{eq:KM2a})-(\ref{eq:KM2c}) and (\ref{eq:Heisenberg}) given by two copies of the $U(1)$ current algebra and the Heisenberg algebra for $(k_0, \Pi_0)$.  The holonomy $k_0$ can then be changed by acting with its conjugate, so that the system is not confined to a single Kac-Moody orbit.  The symplectic form is not degenerate when including $k_0$ because its conjugate $\Pi_0$ also appears. The kinetic term of the above action is just the corresponding geometric action.

\subsubsection{Periodicity of $\mu$}

It is  useful to rewrite (\ref{eq:holonomyU(1)}) in terms of the $U(1)$ group element $e^{i \mu}$ as
\be
A_j = \frac{1}{i} e^{-i \mu} \partial_j e^{i \mu} + k_0\, \delta_{j\varphi} \,.
\ee
The group element $e^{i \mu}$ is assumed to be periodic.  This implies that the function $\mu$, in fact, need not be periodic but is only requested, a priori, to change by an integer multiple of $2 \pi$ as one makes a full turn.

It follows that the pairs $(\mu , k_0)$ and $(\mu + m \vp, k_0 - m)$ ($m \in {\mathbb Z}$) determine the same connection $A_i$ (the function $e^{i m \vp}$ is well defined everywhere on the annulus and so the redefinition $e^{i \mu} \rightarrow e^{i \mu} e^{i m \vp}$ is perfectly acceptable).   The holonomy $k_0$ is thus defined up to an integer.  

By using this ambiguity, one can assume $\mu$ to be periodic.  Indeed, if $\mu$ is periodic up to $2 \pi n$, the function $\mu - n \vp$ is periodic. Taking $\mu$ to be strictly periodic (and not periodic up to an integer multiple of $2 \pi$) is therefore not a restriction in the annulus case.

\subsubsection{Gauge fixing}
A good gauge condition should be a condition that freezes only the proper gauge transformations, without affecting the freedom of performing improper gauge transformations.

For that reason, a condition such as  $A_r = 0$ is too strong. That it is incorrect to impose $A_r = 0$ can be seen from many angles.  To reach the condition $A_r = 0$ from a configuration that does not obey it, one needs to add the gradient $\partial_i \epsilon$ of a function $\epsilon$ that must obey a first-order differential equation with respect to $r$ of the form $\partial_r \epsilon = \cdots$, the solution of which will in general not vanish at the boundaries and will thus involve an improper gauge transformation.  It is also clear that $A_r=0$ is a condition incompatible with a non-vanishing radial Wilson line.  

A gauge condition such as $\partial_\vp A_r = 0$ would also be too strong.  By contrast, the condition $\partial_r A_r = 0$ is acceptable since in order to reach it, one needs to perform a gauge transformation that obeys a second-order differential equation with respect to $r$ of the form $\partial^2_r \epsilon = \cdots$, which can consistently be assumed to vanish at the boundaries.  In fact, a proper gauge transformation need not vanish at the boundaries but must reduce to a constant, the same at both boundaries, so that the residual gauge symmetry in the gauge $\partial_r A_r = 0$ is given by $\epsilon = $constant.

In the gauge $\partial_r A_r = 0$, the constraint $F_{r\vp} = 0$ can be integrated as follows:
\begin{itemize}
\item One can give as boundary data the two sets of Kac-Moody currents $j(\vp)$ and $m(\vp)$ as well as the radial Wilson line $\Pi_0$. One gets from the gauge condition $\partial_r A_r = 0$ that $A_r$ depends only on $\vp$, and from the zero curvature condition $\partial_r A_\vp = \partial_\vp A_r$ that 
$ A_\vp(r, \vp) = (r-r_1) \partial_\vp A_r (\vp) + m(\vp) $ so that
$$ \partial_\vp A_r (\vp) = \frac{j(\vp) - m(\vp)}{r_2-r_1}\,. $$
This equation is consistent because the matching condition implies that $j(\vp) - m(\vp)$ has no zero mode.  It determines $A_r$ up to a constant, which is then fixed by the condition that the radial Wilson line should be equal to $\Pi_0$.  The boundary fields $\Phi$ and $\Psi$ are finally determined up to the addition of a constant, which is the residual gauge symmetry in the gauge  $\partial_r A_r = 0$.
\item One can alternatively give ``initial'' (in $r$) data $ m(\vp)$ and $A_r(\vp)$ at the inner boundary and integrate outwards to the outer boundary.   One gets the Kac-Moody current at the outer boundary as $ j(\vp) = (r_2-r_1) \partial_\vp A_r (\vp) + m(\vp) $, and the radial Wilson line as $\Pi_0 = - \frac{k}{2 \pi}(r_2-r_1) \oint d \vp A_r(\vp)$.
\end{itemize} 

Finally, we note that although one cannot impose $A_r = 0$ throughout the annulus, there exist consistent gauge conditions such that $A_r =0$ at both boundaries (or even in the vicinity of the boundaries). Indeed, if $\epsilon$ is the parameter of the gauge transformation needed to implement the gauge condition, one finds that $\epsilon$ and its radial derivative $\partial_r \epsilon$ are fixed at the two boundaries ($\epsilon(r_1) = \epsilon(r_2) = 0$ because it must be a proper gauge transformation and $\partial_r \epsilon(r_i)$ is determined by the condition that $A_r=0$ at the boundaries).  There are clearly many functions that fulfill these rather weak conditions.  For instance, if $A_r = \left(\frac{r-r_1}{r_2 - r_1}\right)^2$, one finds that $\bar{A}_r = A_r + \partial_r F$ with $F= \frac{1}{(r_2 - r_1)^2} (r-r_1)^2 (r_2-r)$, differs from $A_r$ by a proper gauge transformation since $F$ vanishes at $r_1$ and $r_2$ and is equal to
$ \bar{A}_r = \frac{2}{(r_2-r_1)^2} (r-r_1)(r_2-r)$, an expression that vanishes at the boundaries.

\subsection{Extension to the non-abelian case}
The non-abelian case proceeds conceptually along the same general lines \cite{Elitzur:1989nr}.  The orbits of the coadjoint action of the group are again determined by the holonomy (monodromy) up to conjugation.  

\subsubsection*{Solution of (spatial) zero curvature condition}
More explicitly, in the case of the annulus, one solves the zero curvature condition $F_{r \vp} = 0$ (at any given time) as 
\be
A_r = G^{-1} \partial_r G, \qquad A_\vp = G^{-1} \partial_\vp G + G^{-1} K G\,, \label{eq:NADec}
\ee
where $G(r, \vp)$ is a periodic group element and $K$ is a Lie algebra element that can be taken to be independent of $r$ and $\vp$, and which defines the holonomy.  Both $G$ and $K$ depend in general on $t$.  There is some ambiguity in this decomposition since if $R$ is a Lie algebra element that commutes with $K$ such that $e^{2 \pi R} = I$, then $G \sim e^{R \vp} G$ and $K \sim K - R$.  This ambiguity can be used to impose convenient conditions if one so wishes.

As in the abelian case, the decomposition (\ref{eq:NADec}) involves a gauge redundancy, $G \rightarrow \omega G$, $K \rightarrow \omega K \omega^{-1}$, where $\omega(t)$ is an arbitrary group element that depends only on time, and which indeed drops from (\ref{eq:NADec}).

\subsubsection*{Kac-Moody currents}
The Yang-Mills gauge symmetry $A_\mu \rightarrow S^{-1} \partial_\mu S + S^{-1} A_\mu S$ reads $G \rightarrow GS$ and $K \rightarrow K$ and defines proper gauge transformations  when $S$ vanishes at the boundary.  When $S$ does not vanish at the boundary (and has a non-vanishing charge), it defines an improper gauge transformation that generically  changes the physical state of the system.  The corresponding charge-generators are $A^a_\varphi(r=r_2) \equiv \frac{k}{2\pi} j^a$ (outer boundary) or $A^a_\varphi(r=r_1) \equiv \frac{k}{2\pi} m^a$ (inner boundary).  Both $\{j^a(\vp)\}$ and $\{m^a(\vp)\}$ form a Kac-Moody algebra \cite{Witten:1989hf,Moore:1989yh,Elitzur:1989nr}.  They are called boundary Kac-Moody currents.  

The holonomy is dynamical in the annulus case.  Its conjugate momentum is a global degree of freedom that is not contained in the boundary Kac-Moody currents.  It can be related to Wilson lines connecting the two boundaries (see below).

\subsubsection*{Holonomy matching condition}

The effect of the gauge redundancy is to impose a matching condition between the currents at the two boundaries, expressing that the holonomy (which is a functional of the currents), is the same at both boundaries.  This matching condition is a first class constraint, generating the gauge redundancy.

\begin{figure}
 	\centering
 	\includegraphics[width = 0.5\textwidth]{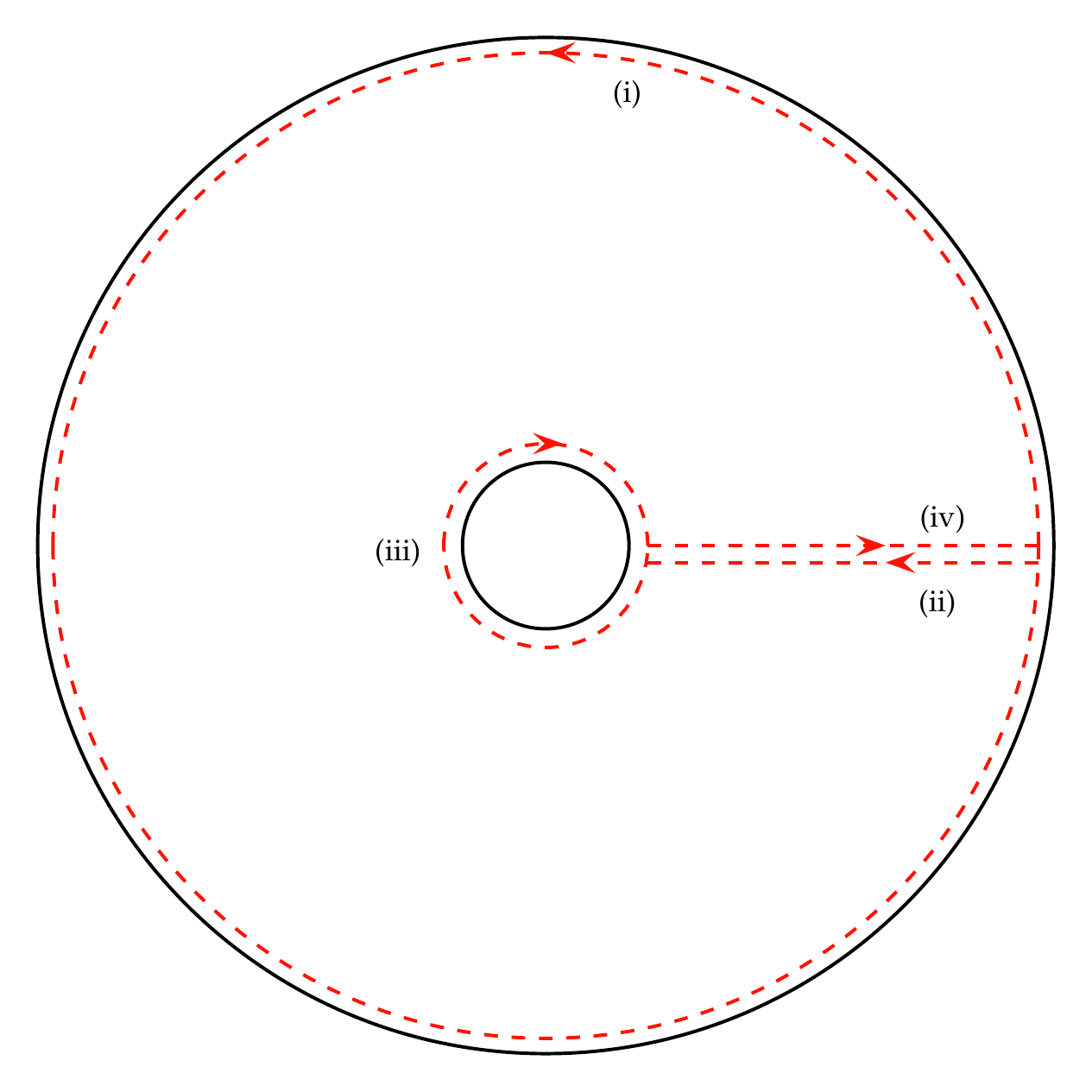}
 	\caption{\label{fig2} A contractible contour composed of (i) the outer boundary, (ii) the radial ray $\rho$  joining the outer boundary to the inner boundary at $\vp = 0$, (iii) the inner boundary (traveled in negative direction), (iv) the ray $\rho$ traveled in the other direction.  
 	}
 \end{figure}

When expressed in terms of the boundary Kac-Moody currents, the matching condition explicitly reads
\be
\cP \exp{\left[-\frac{k}{2\pi}\oint_0 j(\vp) d \vp \right]}= C \, \cP \exp{\left[-\frac{k}{2\pi}\oint_0 m(\vp) d \vp \right]}  \, C^{-1} \, , \label{eq:MatchingNA}
\ee
where the closed integral around the circle has conventionally been taken to start at $\vp = 0$ and where $C$ is the radial path-ordered integral 
\be C \equiv \cP \exp {\left[-\int_{r_1}^{r_2} A_r (r, \vp = 0)dr \right]} = G^{-1}(r=r_2, \vp=0) G(r=r_1, \vp=0)\,.\ee  
This condition expresses the triviality of the holonomy around the contractible contour shown in figure \ref{fig2}.

It is important to realize that the Kac-Moody currents do not fix completely $C$, which contains therefore extra information.  Indeed, $C$ is determined by (\ref{eq:MatchingNA}) up to a transformation that commutes with the holonomy, $C \rightarrow C U$, such that
\be
U \, \cP \exp{\left[-\frac{k}{2\pi}\oint_0 m(\vp) d \vp \right]} \, U^{-1} = \cP \exp{\left[-\frac{k}{2\pi}\oint_0 m(\vp) d \vp \right]}\,.
\ee  
The extra information contained in $C$ is accordingly a specification of $U$.  If the holonomy is put in the Cartan subalgebra, $U$ itself will generically be in the Cartan subalgebra, which means that the amount of extra information contained in $C$ is parametrized by the rank of the group.

\subsubsection*{Radial Wilson lines}

We have just seen that the Wilson lines joining the two boundaries play an important role in the description of the system.  Through the zero curvature condition, any such Wilson line can be expressed in terms of a  radial Wilson line at fixed angle $\bar{\vp}$, $C(\bar{\vp}) \equiv \cP \exp {[- \int_{r_1}^{r_2} A_r (r, \vp = \bar{\vp})dr ]}$ and the Kac-Moody currents at the boundaries.  There is thus only one independent Wilson line connecting the boundaries that must be added to the Kac-Moody currents, to get a complete description of the system. We have conventionally taken $C \equiv C(\bar{\vp}=0)$.
The radial Wilson line $C$ is manifestly gauge invariant under the $\omega$-redundancy.  It transforms as 
$C \rightarrow S^{-1}(2) C S(1)$ under the Kac-Moody symmetries, where $S(2)= S(r=r_2, \vp=0)$ and $S(1)= S(r=r_1, \vp=0)$.  A non-trivial radial Wilson line is again an obstruction to imposing the gauge $A_r= 0$.

The radial Wilson line $C$ has a non-vanishing Poisson bracket with the holonomy (see \cite{Nelson:1989zd,Nelson:1990ba,Nelson:1991an} for the explicit evaluation in $2+1$ gravity).  Including it gives a non-degenerate symplectic structure, which is otherwise degenerate since the Casimirs of the group - in number equal to the rank - do not have a conjugate momentum if we only consider the Kac-Moody currents. This will be illustrated in the gravity case below. However, it should be stressed that the Poisson brackets of the Wilson lines with the other canonical variables do not take in general the simple canonical form that we found in the abelian case, i.e., the radial Wilson lines do not coincide with the canonically conjugate momentum to the holonomy, but are  more complicated phase space functions. To get Darboux coordinates necessitates an appropriate and in general rather cumbersome redefinitions of the variables.

\subsubsection*{Time evolution}
Finally, the time evolution of the fields is given by the temporal components of the zero curvature condition.  It takes the form of a Yang-Mills gauge transformation with gauge parameter equal to $A_t$. Specifying the dynamics requires therefore a choice of $A_t$.  Different boundary values of $A_t$  define different physical Hamiltonians since the Yang-Mills gauge symmetry is improper at the boundary. There is a great freedom in the choice of $A_t$. The choices $A_t = \pm A_\vp$ turns out to be relevant for gravity \cite{Coussaert:1995zp}.

\subsubsection*{Conclusion}
A complete set of observables is given by the Kac-Moody currents at the two boundaries and the conjugate to the holonomy.  The Kac-Moody currents are subject to the matching condition (\ref{eq:MatchingNA}).  This condition takes a more complicated form than in the abelian case (where it reduces to the simple condition $j_0 = m_0$) because the group elements do not commute.  Another way to understand this more intricate form comes from the dynamical role played by the matching condition, which generates the  gauge redundancy parametrized by $\omega(t)$.  In the abelian case, the $\omega$ gauge redundancy  simply coincided with a particular choice of the Yang-Mills gauge symmetry, namely $S = \omega(t)$.  This is not true anymore in the non-abelian case.  For that reason, the generator of the $\omega$-symmetry takes a more involved form when expressed in terms of the generators of the Kac-Moody currents (and the Wilson lines along radial rays).

\section{Pure gravity}\label{sec:gravity}

We now turn to $AdS_3$ gravity, which can  be viewed as a Chern-Simons model with gauge group $SL(2,\mathbb{R}) \times SL(2,\mathbb{R})$ \cite{Achucarro:1987vz,Witten:1988hc}.  The black hole was derived in \cite{Banados:1992wn} and its global structure (Kruskal coordinates and Penrose diagrams) was analyzed in \cite{Banados:1992gq}.

In the eternal black hole case, there are two asymptotic regions and the spatial sections have the topology of a cylinder with a ``radial coordinate'' (e.g., Kruskal coordinate $u$) ranging from $- \infty$ to $+ \infty$ (the two asymptotic regions are connected by a ``wormhole'').  By a gauge transformation, one can eliminate the radial dependence from the $SL(2,\mathbb{R})$ connection in the asymptotic regions (but not everywhere as we have seen)
so that the problem can be effectively reformulated on a cylinder with finite ``height'', or equivalently, on the annulus (\cite{Coussaert:1995zp} and appendix \ref{app:cylinder}) .  This is precisely the context of the previous section.

 We consider thus from the outset the case of a spacetime manifold $\cM_3$ that has the topology of the annulus $\times$ time.  The analysis proceeds as above, but there is an extra feature, namely, that the $AdS_3$ boundary conditions imply a ``Hamiltonian reduction'' at the boundary \cite{Coussaert:1995zp} (along Drinfeld-Sokolov lines \cite{Drinfeld:1984qv,Alekseev:1988ce,Bershadsky:1989mf,Forgacs:1989ac,Feher:1992yx,ORaifeartaigh:1998pox}), leading to the $AdS_3$ asymptotic Virasoro algebra found in \cite{Brown:1986nw}.

As emphasized in the appendix of \cite{Henneaux:1999ib}, in order to deal with the constraints imposed by the $AdS_3$ Hamiltonian reduction, one can either treat independently the two factor subgroups $SL(2,\mathbb{R})$, corresponding to opposite chiralities,  or one can combine them prior to enforcing the constraints in order to get the non-chiral Liouville model.  This latter method was the one initially followed in \cite{Coussaert:1995zp}, but turns out to be cumbersome for dealing with the zero modes.  Accordingly, it is more convenient not to recombine the chiralities.  This is the approach that will be adopted here.

\subsection{WZWN Action }
The two boundaries of the annulus, $\Sigma_o$ for the outer boundary and $\Sigma_i$ for the inner boundary, are the two asymptotic boundaries of the eternal black hole and we will equip as above spacetime with a coordinate system $(r,t,\vp)$ of orientation $\ve_{rt\vp} =1$.  As we have seen, the holonomy is dynamical for the annulus topology.  This means that it is not fixed to some definite value.  In terms of the  black hole charges, we allow accordingly in our phase space black holes with different mass and angular momentum.

The Hamiltonian form of the Chern-Simons action
\begin{equation}\label{Scs}
S_{\textsc{cs}}[A] =  \frac{k}{4\pi} \int_{\cM} \tr\left( A \wedge d A + \frac23 A \wedge A \wedge A \right)\,,
\end{equation}
 is a direct generalization of the abelian one and reads
\begin{equation}\label{ScsHam}
S[A] = \frac{k}{4\pi} \int_{\cM} dt  d\varphi dr\, \tr \left( A_{\varphi} \dot{A}_{r} - A_{r} \dot{A}_{\varphi} + 2 A_{t} F_{r \varphi} \right) + I_{\Sigma_{i}} + I_{\Sigma_{o}}\,,
\end{equation}
with
\begin{equation}
\label{Fphir}
F_{r \vp} = \partial_{r} A_\vp - \partial_\vp A_r + [A_r, A_\vp]\,.
\end{equation}
and where  $I_{\Sigma_{i,o}}$ are boundary terms adapted to the boundary conditions under consideration.

We impose the boundary condition $A_- = 0 $ of \cite{Coussaert:1995zp} at the outer boundary. 
We also choose for definiteness the Hamiltonians on the respective boundaries to have the same sign, which one could interpret as having time evolution on both sides of the black hole run in the same direction. This is achieved by taking  $A_+ = 0$ at the inner boundary.  These choices lead to the boundary terms:
\begin{equation}
\label{Hamiltonian}
I_{\Sigma_{i,o}} = -\frac{k}{4\pi} \int_{\Sigma_{i,o}} dt d\vp \, \tr A_\vp^2\,.
\end{equation}

The next step is to solve the constraint $F_{r\vp} = 0$.   Given that the spatial sections have the annulus topology, they can support non-trivial holonomies along the non-contractible closed loops around the hole. There are two ways to take the holonomy into account. One is just the approach adopted previously, in which group elements $G$ are requested to be periodic and the holonomy appears explicitly,
\begin{equation}\label{explicitAphi}
A_{\vp} = G^{-1}(\partial_\vp + K(t)) G\,, \qquad G(\vp+2\pi) = G(\vp).
\end{equation}
Here, $K(t)$ is a Lie algebra valued function of time that parametrizes the holonomy.\footnote{\label{footnote3} This choice is actually a restriction, as it is only possible to eliminate the $\vp$-dependence in $K(t)$ for simply connected groups, which $SL(2,\mathbb{R})$ is not.}

Another approach is to include the holonomy in the periodicity of the group element. Denoting non-periodic group elements and functions by using tildes, we see that we may equivalently take:
\begin{equation}\label{implicitAphi}
A_\vp = \tilde{G}^{-1}  \partial_{\vp} \tilde{G} \,, \qquad \tilde G (\vp+2\pi) = \exp(2\pi K(t)) \tilde G(\vp)\,.
\end{equation}
Using this it is clear that the relation between $G$ and $\tilde G$ is 
\begin{equation}
\label{GtildeG}
\tilde G = e^{K(t) \vp} G\,.
\end{equation}
We will choose to take periodic group elements and represent the holonomy explicitly in the action. We have verified that the other approach gives the same result, but for the sake of brevity we will not carry out both here.\footnote{In the case of including the holonomy in the periodicity of the group element, one would have to keep track of $\vp$-boundary terms. These can be dealt with using the periodicity conditions \eqref{implicitAphi}, which reduces them to a total $r$-derivative, leading to a non-trivial contribution at the $r$-boundary.}

The action with explicit holonomy can be obtained by substituting \eqref{explicitAphi} into \eqref{ScsHam}. 
The result is (formula (A.7) of  \cite{Henneaux:1999ib})
\begin{align}\label{Sexpl}
S_{\rm CS}[G, K(t)] = & + \frac{k}{4\pi} \int_{\cM} d^3x \; \tr \left( \partial_r (  G^{-1} \partial_\vp  G  G^{-1} \partial_t  G ) \right) + \frac{k}{12\pi} \int_{\cM}  \; \tr ( G^{-1} d  G)^3 \\
& + \frac{k}{4\pi} \int_{\cM} d^3x \; \tr \left(2 \partial_r ( G^{-1} K \partial_t G ) - \partial_t (G^{-1} K \partial_r G) \right)\nonumber + I_{\Sigma_{i}} + I_{\Sigma_{o}}\,,
\end{align}
Here we have discarded a total $\vp$-derivative, which is allowed since $G$ is periodic in $\vp$. In addition, we have also dropped boundary contributions at the time boundaries, and we will continue to do so in the sequel, up to the point where we discuss them systematically. The reason that we delay the discussion of the boundary terms at the time boundaries  is that their form depends on what is kept fixed there, i.e.,  with which representation one is dealing (which complete set of commuting observables is fixed at the time boundaries).  It is premature to discuss them already now without a better grasp of the structure of the physical phase space.

The action now decomposes into two boundary contributions which are coupled through the holonomy parameterized by $K$.
\begin{align}\label{WZW2bdy}
S_{\rm CS}[h,l, K(t)]  = & +  \frac{k}{4\pi} \int_{\Sigma_o} dt d\vp \; \tr \left( h^{-1} \partial_\vp  h  h^{-1} \partial_-  h  + 2 h^{-1} K \partial_- h   - K^2 \right)  \\
& - \frac{k}{4\pi} \int_{\Sigma_i} dt d\vp \; \tr \left( l^{-1} \partial_\vp  l  l^{-1} \partial_+  l  + 2 l^{-1} K \partial_+ l   + K^2 \right)  \nonumber + I_{WZ}[G] \,, \nonumber
\end{align}
where here:
\be
h= G(t, r=r^{\textrm{outer}}, \varphi), \qquad l = G(t, r=r^{\textrm{inner}}, \varphi)\,.
\ee
The Wess-Zumino term
\begin{equation}
I_{WZ}[G] =  \frac{k}{12\pi} \int_{\cM}  \; \tr ( G^{-1} d  G)^3  \,,
\end{equation}
can be written as a total derivative and hence it also only depends on the boundary values of the group element $G$.

The action \eqref{WZW2bdy} is invariant under the gauge symmetry $G \rightarrow \omega(t) G$, $K \rightarrow \omega(t) K \omega^{-1}(t)$, which implies in terms of the boundary fields,
\be \label{gaugetrafo}
h \rightarrow \omega(t) h, \qquad l \rightarrow \omega(t) l, \qquad K \rightarrow \omega(t) K \omega^{-1}(t)\,.
\ee
This gauge invariance results from the redundancy of the parametrization of the group element $G$ \cite{Elitzur:1989nr} and can  straightforwardly be verified to be present in the above action\footnote{In that respect, the comments made in the literature that the gauge symmetry would have allegedly been overlooked in \cite{Coussaert:1995zp,Henneaux:1999ib} make us somewhat perplexed since this gauge symmetry is manifestly present in the action.}.

\subsection{More on the holonomy}

With the boundary condition $A_t =  A_\vp$, the equation of motion $F_{t \vp}=0$ implies $\partial_- A_\vp = 0$ at the outer boundary.  Similarly, one gets $\partial_+ A_\vp = 0$ at the inner boundary.  

The equation $\partial_- A_\vp = 0$ at the outer boundary reads explicitly $ \partial_- (h^{-1} \partial_\vp h) + \partial_- (h^{-1} K h) = 0$
and can be rewritten as
\be 
h^{-1} \partial_\vp ( \partial_-h h^{-1})h + \partial_- (h^{-1} K h) = 0,
\ee
or
\be
\dot{K} = - \partial_\vp (\partial_- h h^{-1}) + \partial_-h h^{-1} K - K \partial_-h h^{-1} .
\ee
Integrating over $\vp$ yields
\be
\dot{K} = [a, K] \,,  \label{eq:EOMK}
\ee
with
\be
a = \frac{1}{2 \pi} \oint  \partial_-h h^{-1} d \vp,
\ee
an equation that makes sense since $a$ transforms as a connection for the gauge transformations (\ref{gaugetrafo}),   
\be
a \rightarrow \dot{\omega} \omega^{-1} + \omega a \omega^{-1} \,,
\ee
so that
\be 
D^{(a)}_t K \equiv \dot{K} - [a,K]  \rightarrow \omega (D^{(a)}_t K) \omega^{-1}.
\ee
Although transforming simply as a connection for the gauge transformations (\ref{gaugetrafo}), the  
transformation law of $a$ is intricate under the Kac-Moody symmetry. 

The equation (\ref{eq:EOMK}) could of course have been derived directly from the action \eqref{WZW2bdy} and is equivalent to the similar equation obtained at the inner boundary thanks to the zero curvature condition $F_{r \vp} = 0$.  Its solution is
\be
K(t) = S(t) K(0) S^{-1}(t), \qquad S(t) = T \exp \int_0^t a(\tau) d \tau, 
\ee
which shows that under time evolution, the holonomy of the bulk Chern-Simons connection $K$ stays in the same conjugacy class.

This holonomy  is so far an arbitrary element of the $SL(2,\mathbb{R})$ algebra. Due to the gauge symmetry \eqref{gaugetrafo} we see that  only the conjugacy class of $K$ has any physical relevance, and furthermore, we just proved that this conjugacy class is constant in time. By a gauge transformation, we may always put $K$ into a form where it is given by a ``canonical'' element of either one of the three conjugacy classes of $SL(2,\mathbb{R})$
\begin{itemize}
	\item[] {\bf Hyperbolic:} Conjugate to an element $K(t) = k_0(t) L_0$ \,,
	\item[] {\bf Elliptic:} Conjugate to an element $K(t) = \frac12 k_e(t)  (L_- - L_+)$\,,
	\item[] {\bf Parabolic:} Conjugate to an element $K(t) = k_p(t) L_+$\,. One can in fact set $k_p = 1$ by redefinitions, but we keep a general $k_p$ to check that it does indeed drops from the final form of the action. 
\end{itemize}
where $L_n$ with $n = -,0,+$, are generators of $SL(2,\bR)$ algebra satisfying the algebra \eqref{bracket-init}.

We will now analyze these three possibilities independently and thereby cover all cases. This is technically more tractable than keeping $K$ arbitrary.

\subsection{Hyperbolic holonomy}\label{sec:hyperbolic}

We will first study the case of hyperbolic holonomy, i.e., we choose
\begin{equation}
K(t) = k_0(t) L_0 \,, \qquad k_0 \not = 0 \, .
\end{equation}
 This choice not only forces the holonomy to be hyperbolic, but also partly freezes the gauge freedom (\ref{gaugetrafo}), since $K(t)$ will not remain diagonal under arbitrary conjugation.  The gauge transformations that preserve the form of $K(t)$ must be such that 
 \be
 \omega(t) (k_0 L_0) \omega^{-1}(t) = k'_0 L_0 \,,
 \ee
and this imposes 
\begin{equation}
\omega(t) = e^{\lambda_0(t)L_0}\,.
\end{equation}
The residual gauge symmetry is abelian and non-compact, i.e., parametrized by $\mathbb{R}$. 

One can insert the gauge condition $K(t) \in \{\xi L_0 \}$ inside the action because this is a ``canonical gauge'' \cite{Henneaux:1992ig}.   In that gauge, the holonomy $K(t) = k_0(t) \, L_0 $ is  constant, $\dot{k}_0 = 0$.

To continue, we decompose the dynamical matrix $G$ in terms of simpler matrices.  A popular choice is the Gauss decomposition. Its advantage is that it is convenient for highest-weight representations of $SL(2,\bR)$ and suitable for hyperbolic holonomies. The disadvantage of the Gauss decomposition is that it is not globally accessible, i.e. not all $SL(2,\bR)$ elements have a Gauss decomposition \cite{Tsutsui:1994pp}. 

The group elements $h$ and $l$ are parameterized as
\be \label{Gauss}
h = e^{Y L_-} e^{ \Phi L_0} e^{X L_+}, \qquad l = e^{\tilde{Y} L_-} e^{ \tilde{\Phi} L_0} e^{\tilde{X} L_+} \,.
\ee
The group element $G$ depends on all coordinates, but its pullback to the $r$-boundaries is, of course, $r$ independent. The functions $Y, \Phi, X$ and $\tilde{Y}, \tilde{\Phi}, \tilde{X}$ depend on the boundary coordinates $t$ and $\vp$. As we will be interested in imposing lowest-weight boundary conditions at the inner boundary, it is much more convenient to work with a parametrization of $l$ at $r=r_1$ such that
\be \label{Gaussinner}
l = e^{V L_+} e^{\Psi L_0} e^{U L_-}.
\ee
This can be achieved by following field redefinitions at the inner boundary:
\bea
\tilde{\Phi} &=& \Psi + 2 \log (1 + e^{-\Psi} UV),\\
\tilde{Y} &=& \frac{e^{- \Psi} U}{(1 + e^{-\Psi} UV)},\\
\tilde{X} &=& \frac{ e^{-\Psi}  V}{(1 + e^{-\Psi} UV)},
\eea
where the functions $V$, $\Psi$, $U$ depend on the boundary coordinates $t$ and $\vp$. We emphasize that this field redefinition is only done at the inner boundary, such that the all $SL(2,\mathbb{R})$ elements are defined using a unique Gauss decomposition, not only at both boundaries, but also in the bulk.

Using the Gauss decomposition, the action \eqref{Sexpl} becomes
\begin{equation}
S_{CS}[G,K] = S_o - S_i + S_{\rm hol} \,,
\end{equation}
with
\begin{align}\label{So}
S_o & =  \frac{k}{4\pi} \int_{\Sigma_o} dt d\vp \left( \frac12 \partial_- \Phi \Phi' + 2 e^\Phi \partial_- X Y' \right)\,, \\
S_i & =  \frac{k}{4\pi} \int_{\Sigma_i} dt d\vp \left( \frac12 \partial_+ \Psi \Psi' + 2 e^{-\Psi} \partial_+ U V' \right)\,,
\end{align}
and
\begin{align}
S_{\rm hol} = & \, \frac{k}{4\pi} \int  dt d\vp\; \left[ k_0 \left(\partial_- \Phi - \partial_+ \Psi  - 2 e^{\Phi} Y \partial_- X - 2 e^{-\Psi} V \partial_+ U \right) -  k_0^2 \right] 
\end{align}
(compare with formula (A.13) of \cite{Henneaux:1999ib}).
Another way to split the action among the two boundaries is to take
\begin{equation}
S_{\rm CS}[k_0, Y,\Phi,X,V,\Psi,U] = S_{\rm bdy}^{\Sigma_o}[k_0, Y, \Phi, X] - S_{\rm bdy}^{\Sigma_i}[k_0, V, \Psi, U] \,,
\end{equation}
with
\bea
S_{\rm bdy}^{\Sigma_o}[k_0, Y,\Phi,X] &=& \frac{k}{4\pi} \int dt d\vp \; \left( \frac{1}{2}\partial_- \Phi (\Phi' + 2 k_0) + 2 e^{\Phi} \partial_- X (Y' - k_0 Y) - \frac{1}{2} k_0^2 \right)\,,\nonumber\\
S_{\rm bdy}^{\Sigma_i}[k_0, V,\Psi,U] &=& \frac{k}{4\pi} \int dt d\vp \; \left( \frac{1}{2}\partial_+ \Psi (\Psi' + 2 k_0) + 2 e^{-\Psi} \partial_+ U (V' + k_0 V) + \frac{1}{2} k_0^2 \right)\,.\nonumber\\\label{Sbdy}
\eea

The Lagrangian is easily checked to be invariant up to total derivative terms under the residual gauge symmetry,
\begin{subequations}
\begin{align}
\Phi \to \hat \Phi & = \Phi + \lambda^0 \, , \\
Y \to \hat Y & =  Y  e^{- \lambda^0}  \, ,\\
X \to \hat X & = X  \,, \\
k_0 \to \hat k_0 & = k_0 \, ,
\end{align}
\end{subequations}
with similar expressions holding for $\Psi, V $ and $U$ with appropriate changes implemented, i.e. if $\Psi \to \hat \Psi = \Psi + \lambda^0$ then in order to compensate for that $V \to \hat V = V e^{\lambda^0}$ .

\subsubsection{Boundary conditions}
Next we impose the reduction conditions on the Chern-Simons connection that express standard asymptotic $AdS_3$ behaviour \cite{Coussaert:1995zp}. We consider explicitly one asymptotic boundary only (the outer boundary).   Similar considerations apply to the inner boundary. The only difference in their treatment is the choice of $SL(2,\mathbb{R})$ representation at each boundary. While the boundary conditions on the fields at outer boundary are in accordance with highest-weight representation, those on the fields at inner boundary are in accordance with the lowest-weight representation. As shown in \cite{Coussaert:1995zp} and discussed in Appendix \ref{app:cylinder}, the boundary conditions on the fields at $r=r_2 \equiv r_o$ are \begin{equation}
A_r = 0, \qquad A_\vp = L_- + \cL(t,\vp) L_+\,.
\label{eq:DSReduction}
\end{equation}
Similarly, the boundary conditions on the fields at $r=r_1 \equiv r_i$ are \begin{equation}
A_r = 0, \qquad A_\vp = L_+ + \cM(t,\vp) L_-\,.
\label{eq:DSReduction_lowest}
\end{equation}
In terms of the field appearing in the Gauss decomposition, this gives the conditions
\begin{subequations}
	\label{Gaussconstraints}
	\begin{align}
	e^{\Phi} (Y' - k_0 Y) & = 1  \,, \\
	\Phi' + k_0 & = 2 X \,, 
	\end{align}
\end{subequations}
and the expression for $\cL$ (compare again with Appendix of \cite{Henneaux:1999ib})
\be
X' + X^2  = \cL \, ,  \label{eq:CalLX}
\ee
with the similar expressions holding for $\Psi$, $V$ and $U$ at the inner boundary. One can find them by the following substitutions $\Phi \to -\Psi$, $k_0 \to -k_0$ and $Y,X,\cL \to V,U,\cM$. 

Since $k_0 \not=0$, one can solve the first condition to express $Y$ in terms of $\Phi$ and $k_0$.  This can be verified by Fourier expansion of $Y$ and $e^{-\Phi}$.  Note that the equation would be inconsistent if $k_0$ were to vanish since one would have then $Y' = e^{-\Phi}$, yielding upon integration over $\vp$ the contradictory statement $0 = \oint e^{- \Phi} d \vp>0$.  Similarly, the second equation enables one to express $X$ in terms of $\Phi$ and $k_0$.

If one inserts the resulting expressions into the action, as it is permissible (see Appendix \ref{app:Consistency} for more information on this point), one gets
\begin{equation}\label{Sred}
S_{\rm bdy}^{\Sigma_o}[k_0, \Phi] =  \frac{k}{4\pi} \int dt d\vp \; \left( \frac{1}{2}\partial_- \Phi (\Phi' + 2 k_0) - \frac12 k_0^2 \right)\,,
\end{equation}
where we have dropped a total derivative term.  
Reinstating the fields on the inner boundary, for which similar steps can be taken,  yields the total action 
\begin{equation}\label{Shyper}
S[k_0, \Phi, \Psi] =  \frac{k}{4\pi} \int dt d\vp \; \left( \frac{1}{2}\partial_- \Phi \Phi' - \frac{1}{2}\partial_+ \Psi \Psi' +  k_0( \partial_- \Phi - \partial_+ \Psi) - k_0^2  \right)\,.
\end{equation}
Note that while the kinetic terms for $\Phi$ and $\Psi$ have opposing sign, their Hamiltonians have the same sign and hence are positive definite for both fields. This action is equivalent to the $U(1)$ case studied in the previous section: two chiral boson actions, coupled to the holonomy.  Once again the action is gauge invariant under a diagonal abelian gauge symmetry acting as \eqref{eq:GaugeSymmAnnulus}, although here the symmetry is not compact (but this does not affect the form of the action). The canonically conjugate momentum to the holonomy is gauge invariant and given by
\begin{equation}
\Pi_0= -\frac{k}{4\pi} \int d \vp \; (\Phi - \Psi) \,.
\end{equation}
One can notice that varying the action with respect to
the conjugate $\Pi_0$ will result in the equation $\dot{k}_0 =0$.

Having obtained the reduced action, one can work out the boundary term at the time boundaries.  Modulo the zero modes, the action is that of chiral bosons \cite{Floreanini:1987as}.  The kinetic term for the zero modes has the standard  $p \dot{q}$ form;  it is well known how the boundary term at the time boundaries depends in that case on the chosen representation ($p \dot{q}$ as such being adapted to the $q$-representation where the $q$'s are fixed at the time boundaries). For the chiral boson, which is self-conjugate,  the boundary term at the time boundaries is discussed in \cite{Henneaux:1987hz}, to which we refer for details. 

The other $SL(2, \mathbb{R})$ leads to an action similar to (\ref{Shyper}) coupled to another, independent holonomy, with its own independent conjugate momentum, so that in the end, we get four chiral actions coupled through two different holonomies.

The action (\ref{Shyper}) contains two chiral bosons of opposite chiralities with a common zero mode.  As shown in the discussion of the $U(1)$ case, these can be combined to yield the action for a free boson on the cylinder.  In turn, this action can be classically mapped on the Liouville action by a B\"acklund transformation as shown in \cite{Gervais:1982nw,DHoker:1982wmk,Braaten:1982fr} (see also \cite{Seiberg:1990eb,Henneaux:1999ib,Barnich:2013yka} in the gravity context). The two $SL(2, \mathbb{R})$ factors would lead to two Liouville models. This procedure works only, however, if the boundary conditions at the two boundaries have opposite chiralities, which was a consequence of having positive Hamiltonians on both boundaries. This is much more natural than the combination of the chiral bosons associated with the different $SL(2, \mathbb{R})$ factors at the same boundary considered in \cite{Coussaert:1995zp}, because these do not share the same zero modes.

\subsubsection{Virasoro algebra}

As shown in \cite{Coussaert:1995zp}, the $\cL$'s form the Virasoro algebra of \cite{Brown:1986nw} with central charge $c = \frac{3 \ell}{2G}$. In terms of the Fourier modes of $\cL(\vp)$
\be
[\cL_m, \cL_n] = (m-n) \cL_{m+n} + \frac{c}{12}n (n^2-1) \delta_{m+n,0} \,.
\ee
The classical state of the system at a given time is completely determined (for a single $SL(2, \mathbb{R})$-factor) by giving:
\begin{enumerate}
\item The Virasoro generators $\cL(\vp)$ and $\cM(\vp)$ at the two boundaries, or equivalently, the two $SL(2, \mathbb{R})$ connections $A_\vp$ of the form (\ref{eq:DSReduction}) and \eqref{eq:DSReduction_lowest}. These satisfy the matching conditions (\ref{eq:MatchingNA}), rewritten in (\ref{eq:MatchingGravity}) below.
\item One radial Wilson line along, say, $\vp = 0$,
\be
\label{radialWilson}
C \equiv \cP \exp {\left[-\int_{r_i}^{r_o} A_r (r, \vp = 0)dr \right]} \,,
\ee
where this integral is evaluated in a gauge where $A_r$ vanishes at both boundaries.
\end{enumerate}

The matching condition reads
\be
\cP \exp{\left[-\oint_o \left(L_- + \cL(\vp) L_+\right) d \vp \right]}= C \, \cP \exp{\left[-\oint_i \left(L_+ + \cM(\vp) L_-\right) d \vp\right]}  \, C^{-1}\,. \label{eq:MatchingGravity}
\ee

The physical phase space is spanned by $(\cL(\vp), \cM(\vp), C)$ subject to (\ref{eq:MatchingGravity}).  The Poisson bracket structure in this space is non-degenerate.  Equivalently, one can describe the physical phase space in terms of the chiral bosons $\Phi(\vp), \Psi(\vp)$,  the holonomy $k_0$ and its conjugate $\Pi_0$.  There is a redundancy in the description, since the zero mode of $\Phi(\vp) + \Psi (\vp)$ is pure gauge and its conjugate momentum constrained to vanish.

Given $(\cL(\vp), \cM(\vp), C)$, one determines $(\Phi(\vp), \Psi(\vp), k_0, \Pi_0)$ as follows.  First, one determines $X$ from $\cL$ from (\ref{eq:CalLX}).  As explained in Appendix \ref{app:Consistency}, the solution is unique if one requests $X$ to be a periodic function on the circle. Knowing $X$, one can then determine $k_0$ by integrating the second equation (\ref{Gaussconstraints}) over $\vp$, getting $\pi k_0 = \int d \vp X$. A similar procedure can be applied at the other boundary to yield the corresponding function $U$.  The matching condition guarantees that one gets the same $k_0$ from $U$ as from $X$. Once $k_0$ is known, one determines  $\Phi$ through the second equation (\ref{Gaussconstraints})  (and $\Psi$ through the corresponding equation at the other boundary).  These equations leave arbitrary the zero modes of $\Phi$ and $\Psi$, but the difference of these zero modes is fixed by $C$ (see below).  Only the non-gauge invariant sum is arbitrary, as it should. 

A beautiful formula giving the metric in the bulk in the vicinity of one boundary in terms of the Virasoro generators at that boundary has been derived in \cite{Banados:1998gg}.  It provides the inward integration of the constraints starting from the boundary,  in a gauge where $g_{rr}$ is fixed to be a constant (equivalent to fixing $A_r$ to some prescribed value and valid in the neighbourhood of the boundary).

In order to evolve the fields, one needs to choose $A_t$.  The choices relevant to standard anti-de Sitter asymptotics are, as we have seen, $A_- = 0$ or $A_+ = 0$ \cite{Coussaert:1995zp}.  This leads to a simple time evolution at the boundary, described by the chiral (or anti-chiral) Hamiltonian of (\ref{Shyper}), and yielding chiral (or anti-chiral) Virasoro generators $\cL(x^+)$ (or $\cM(x^-)$).  Other choices of $A_t$, leading to different  asymptotic dynamics with interesting integrable structures, have been studied in
\cite{Perez:2016vqo,Gonzalez:2018jgp,Fuentealba:2017omf}.

\subsubsection{The zero angular momentum black hole}

We illustrate the previous derivation in the case of the eternal zero angular momentum black hole \cite{Banados:1992wn} of mass $M$, the global structure of which was elucidated in \cite{Banados:1992gq}.  In Rindler-like coordinates, one of the $SL(2,\mathbb{R})$ connections reads, on the $t=0$ slice \cite{Cotler:2018zff}
\be
A= -\frac{2}{1-z^2} L_0 dz + \frac{1+z}{2(1-z)} \sqrt{M} L_+ d \vp+ \frac{1-z}{2(1+z)} \sqrt{M} L_- d \vp \,,   \label{eq:ConnectionStatic}
\ee
the asymptotic regions being at $z = \pm 1$ and our conventions for the $SL(2,\mathbb{R})$ generators are listed in appendix \ref{app:conventions}. A similar expression holds for the other $SL(2, \mathbb{R})$ connection and we focus for that reason only on (\ref{eq:ConnectionStatic}). 

The connection is singular at the boundaries, and can be made regular by a gauge transformation  of the type considered in \cite{Coussaert:1995zp,Henneaux:1999ib} at each boundary.  In order to implement this procedure, we proceed in two steps.  First, we regularize the connection everywhere by the gauge transformation 
\be
B = \begin{pmatrix} 
\left(\frac{1+z}{1-z}\right)^{\frac12} & 0 \\
0 & \left(\frac{1-z}{1+z}\right)^{\frac12}
\end{pmatrix} \,,
\ee
such that the new expression  $A' = B^{-1} A B + B^{-1} dB$ for the connection reads
\be
A' = \frac12 \sqrt{M} \left(L_+ + L_- \right) d \vp \,.
\ee
This transformation, however, is not acceptable and must be corrected because it fails to implement the highest (lowest) weight form of the connection at the boundaries.  We thus perform now the gauge transformation with group element 
\be
b = \begin{pmatrix} 
F & 0 \\
0 & F^{-1}
\end{pmatrix} \,,
\ee
which brings the connection to the form 
\be 
A_\vp(z = 1) = L_- + \frac{M}{4} L_+  \,,
\ee at the outer boundary and 
\be
A_\vp(z = -1) = L_+ + \frac{M}{4} L_- \,,
\ee
 at the lower boundary provided the function $F(z)$ is taken to fulfill
 \be
 F' (z=1) = 0 = F'(z=-1), \qquad F^{-2}(z=1) = \frac{\sqrt{M}}{2} = F^2(z=-1) \,.
  \ee
 This leads for the radial Wilson line
 \be
 C = \cP \exp \left[-\int_{z=1}^{z=-1} dz (b^{-1} \partial_z b)\right] = b^{-1} (z=-1) b(z=1) = \begin{pmatrix} 
 \frac{\sqrt{M}}{2} & 0 \\
0 & \frac{2}{\sqrt{M}}
\end{pmatrix} \label{eq:RadialStatic} \,.
 \ee
 This non-trivial radial Wilson line arises because of the incompatibility of imposing simultaneously the radial gauge $A_z=0$ everywhere and the Hamiltonian reduction boundary conditions at  $z = \pm 1$.  We see again very clearly the fact that the gauge condition $A_z=0$ is globally not appropriate.

 We now derive the chiral fields from the boundary Virasoro algebras $\cL(\vp) = \frac{M}{4}$, $\cM(\vp) = \frac{M}{4}$ and the radial Wilson line (\ref{eq:RadialStatic}).   It is obvious that these fields reduce to their zero modes. Since the eigenvalues of the matrices $L_- + \frac{M}{4}L_+$ and $L_+ + \frac{M}{4}L_-$, which coincide, are equal to $\pm \frac{\sqrt{M}}{2}$, we conclude that the holonomy is equal to 
 \be
 k_0 = \sqrt{M} \,,
 \ee
 (recall that  $\begin{pmatrix} 
 1 & 0 \\
0 & -1
\end{pmatrix} = 2L_0$ and that the holonomy in the diagonal gauge reads $k_0 L_0$).
This is in agreement with the direct computation from $X$, which yields $X^2 = \frac{M}{4}$ i.e., $X = \frac{\sqrt{M}}{2}$.  Plugging this into the equation for $\Phi$, one gets the above value of $k_0$.  

The zero modes $\Phi_0$ and $\Psi_0$ are not fixed by the Hamiltonian reduction constraints.  While the sum is pure gauge, one determines  the difference by relating it to the radial Wilson line $C$ and using the Gauss decomposition.   This follows from the relation 
\be
 A_\vp (z=1)= C A_\vp (z=-1) C^{-1} \,,
\ee
which implies $hC = l$. 
In terms of the constrained zero-modes of the fields appearing in the Gauss decomposition \eqref{Gauss} and \eqref{Gaussinner}, this implies
\bea
V_0 &=& \frac{e^{\Phi_0} X_0}{1+ e^{\Phi_0} X_0 Y_0}= \sqrt{M} e^{\Phi_0},\nonumber\\
\Psi_0 &=& \Phi_0 -2  \ln \frac{2}{\sqrt{M}}(1+ e^{\Phi}X_0 Y_0) = \Phi_0 + \ln M,\nonumber\\
U_0 &=& \frac{M}{4} \frac{e^{\Phi_0} Y_0}{1+ e^{\Phi_0} X_0 Y_0} = -\frac{\sqrt{M}}{2}.
\eea
and the difference $\Psi_0 - \Phi_0$ is indeed fixed by $C$.

It is important to emphasize that the non-trivial holonomy, which can be viewed as an obstruction to the gauge $A_z=0$, arises because the boundary conditions at two boundaries are ``twisted'' with respect to one another; highest weight boundary condition at the outer boundary and lowest weight boundary condition at the inner one.

\subsubsection{Relation to the geometric action}

As in the $U(1)$ example, the action on a single boundary is related to a geometric action, but now of the coadjoint orbits of the Virasoro group. That it must be so follows from the fact that the physics of the system is completely captured by the Virasoro generators at each boundary, which transform in the coadjoint representation of the Virasoro group (plus the global mode $C$). The dynamics of the system is just the dynamics of the Virasoro algebras and of $C$. The brackets of the Virasoro generators that follow from the action reproduce the Virasoro algebra, indicating that the symplectic structure coincides  with the natural symplectic structure on the coadjoint orbits.

To exhibit explicitly the connection with the geometric action, we recall that boundary diffeomorphisms appear as residual gauge symmetries compatible with the Hamiltonian reduction constraints, and that these are naturally parametrized in the Gauss decomposition by the $L_-$-factor of the group gauge parameter.  It is therefore no surprise that it turns out to be convenient
 to express the action and the constraints in terms of a field $f(t,\vp)$ directly related to $Y$ (the field associated with the same generator $L_-$), defined as:
\begin{equation}
Y(t,\vp) = \exp\left(- k_0(t)(f(t,\vp) - \vp) \right) \,.
\end{equation}
The periodicity of $f$ is
\be
f(t,\vp +2\pi) = f(t,\vp) + 2\pi \,.
\ee
In terms of this field, the constraints \eqref{Gaussconstraints} become
\begin{equation}
- k_0(t) f'(t,\vp) e^{-k_0 ( f(t,\vp)-\vp)} = e^{-\phi(t,\vp)}\,, \qquad X(t,\vp) = \frac12 k_0(t) f' - \frac{f''}{2 f'}\,,
\end{equation}
and the function $\cL$ becomes
\begin{equation}
\cL = \frac{k_0(t)}{4} f'{}^2 - \frac12 \{f, \vp \}\,,
\end{equation}
where $\{f,\vp\}$ is the Schwarzian derivative of $f$
\begin{equation}
\{f,\vp\} = \frac{f'''}{f'} - \frac{3}{2}\left( \frac{f''}{f'} \right)^2 \,.
\end{equation}
In terms of $f(t,\vp)$ the action \eqref{Sred} becomes (up to total derivatives):
\begin{equation}\label{Sgeom}
S_{\rm bdy}^{\Sigma_o} =  \frac{k}{8\pi} \int_{\Sigma_o} dt d\vp \; \left[ \frac{\partial_- f' f''}{f'{}^2} + k_0(t)^2 \partial_- f f' \right]\,.
\end{equation}
The field equations of this action are explicitly
\begin{equation}
\frac{1}{f'} \partial_- \left( \{ f, \vp \} - \frac12 k_0^2(t) f'{}^2 \right) = 0 \,,
\end{equation}
which are proportional to $\partial_- \cL = 0$.

The action \eqref{Sgeom} is exactly the geometric action on the coadjoint orbits of the Virasoro group as found by the authors of \cite{Alekseev:1988ce}. The Virasoro orbits are  here $\textrm{Diff}(S^1)/S^1$ (also denoted $\textrm{Diff}(S^1)/\mathbb{R}^1$ since the stability subgroup is non compact, see \cite{Khesin} page 79). The relationship between the Virasoro central charge $c$, the orbit representatives $b_0$ and $k$ and $k_0(t)$ is 
\begin{equation}
c = 6k\,, \qquad b_0 = \frac{c}{ 48 \pi} k_0(t)^2\,.
\end{equation}
Coadjoint orbits of the Virasoro group with positive representatives have a $U(1)$ little group \cite{Witten:1987ty}, reflecting the $U(1)$ gauge invariance of the chiral boson action. The holonomy is dynamical, so that one can contemplate initial data not restricted to a single orbit (but, for any chosen set of initial data, the evolution stays on the orbit selected by these initial data since $\dot{k}_0 = 0$).   There is a similar geometric action at the other boundary, with the same value for the orbit representative.

In general the action \eqref{Sgeom}  of \cite{Alekseev:1988ce} can be defined for any real orbit representative $b_0$, but here this quantity is strictly positive. To obtain Alekseev-Shatashvili action for negative orbit representatives, we have to consider bulk holonomies in the elliptic conjugacy class of $SL(2,\mathbb{R})$.

\subsection{Elliptic holonomy}\label{sec:elliptic}

We will now consider the holonomy to be an element in the elliptic conjugacy class of $SL(2, \bR)$. Solutions with elliptic holonomies correspond to point particle sources and were in fact the first to be studied \cite{Deser:1984dr}.  They define conical singularities, of particular interest when the excess angle is a multiple of $2 \pi$ \cite{Balasubramanian:2000rt,Castro:2011iw,Raeymaekers:2014kea}.

We choose the holonomy to have the form $e^{2 \pi K(t)}$ with
\begin{equation}\label{elliptic}
K(t) = \frac{k_e(t)}{2} (L_- - L_+)\,.
\end{equation}

In order to derive the action, it is convenient to use the Iwasawa decomposition to parameterize the group element (see e.g. \cite{Helgason}). This is because the compact subgroup is put in the limelight.  All group elements $g \in SL(2,\bR)$ can be parametrized as the product of an element of the compact subgroup $k$, a diagonal group element $a$ and a nilpotent element $n$. Unlike the Gauss decomposition, the Iwasawa decomposition does hold globally.

We will again take $G(r = r^{\rm outer}) = h$ and $G(r= r^{\rm inner}) = l $ as before, but as we just explained, we now decompose $h$ as
\begin{equation}\label{Iwasawa}
h = k.a.n^+  \,,
\end{equation}
with
\begin{subequations}
\begin{align}
k & = \left(\begin{array}{lr}
\cos \theta(t,\vp) & - \sin \theta(t,\vp) \\
\sin \theta(t,\vp) & \cos \theta(t,\vp)
\end{array} \right) =  \exp( \theta(t,\vp) \left(L_- - L_+ \right) )\,, \\
a & =  \left(\begin{array}{cc}
e^{\Phi(t,\vp)} & 0 \\
0 & e^{-\Phi(t,\vp)}
\end{array} \right) = \exp( \Phi(t,\vp)L_0 )\,, \\
n^+ & =  \left(\begin{array}{cc}
1 & \eta(t,\vp) \\
0 & 1
\end{array} \right) = \exp (\eta(t,\vp) L_+ )\,.
\end{align}
\end{subequations}
For $l$ we can find appropriate field redefinitions at inner boundary such that $l = k.a.n^- $ with $\theta \to \vartheta$, $\Phi \to \Psi$ and 
\be
n^-  =  \left(\begin{array}{cc}
1 & 0 \\
\nu(t,\vp) & 1
\end{array}\right) = \exp (\nu(t,\vp) L_- )\,.
\ee

The matrices $k$, $a$ and $n^{\pm}$ are periodic in $\vp$.  This implies that the fields $\Phi$ and $\eta$ are periodic, but $\theta$ can change by an integer multiple of $2 \pi$ as one goes around the annulus.  By using the ambiguity  described below Eq. (\ref{eq:NADec}) if necessary, one can assume, however, $\theta$ to be strictly periodic.  

As it is well known, in the spinor representation of $SO(2,1)$ used here (i.e., $SL(2, \mathbb{R})$-matrices), anti-de Sitter has $k_e =1$, corresponding to the holonomy $e^{2\pi K} = -I$.  Just as in the hyperbolic holonomy case, there is no solution with $k_e = 0$ as we shall see.

Inserting the Iwasawa decomposition in the action \eqref{WZW2bdy}, and using also \eqref{elliptic},   leads to 
the action
\begin{equation}
S = S^{\Sigma_o}_{\rm bdy}[\theta,\Phi,\eta,k_e] - S^{\Sigma_i}_{\rm bdy}[\vartheta,\Psi,\nu,k_e] \,,
\end{equation}
with
\begin{equation}\label{Siwa}
S^{\Sigma_o}_{\rm bdy} =  \frac{k}{4\pi} \int_{\Sigma_o} dt d\vp \; \left\{\frac12 \partial_- \Phi \Phi' - 2 \partial_- \theta(\theta' + k_e) + 2e^{\Phi} \partial_- \eta \left(\theta' + \frac12 k_e \right) + \frac12 k_e^2 \right\},
\end{equation}
(up to a total time derivative) and
\begin{equation}\label{Siwainner}
S^{\Sigma_i}_{\rm bdy} =  \frac{k}{4\pi} \int_{\Sigma_i} dt d\vp \; \left\{\frac12 \partial_+ \Psi \Psi' - 2 \partial_+ \vartheta(\vartheta' + k_e) - 2e^{-\Psi} \partial_+ \nu \left(\vartheta' + \frac12 k_e \right) - \frac12 k_e^2 \right\}.
\end{equation}

\subsubsection{Boundary conditions and reduced action}
The constraints imposed by the boundary conditions on the outer boundary are now
\begin{align}\label{constraintell}
e^{\Phi} \left(\theta' + \frac12 k_e(t) \right) = 1 \, , && 
\eta = \frac12 \Phi'\,,
\end{align}
and
\begin{equation}\label{Lell}
\cL = - e^{-2\Phi} + \frac14 \left( \Phi'^2 + 2 \Phi'' \right)\,.
\end{equation}
There are similar expressions for the constraints on the inner boundary fields with  $\cL \to \cM$, $\theta \to - \vartheta$, $\Phi \to -\Psi$, $\eta \to \nu$ and $k_e \to - k_e$.  
Again, the first constraint with $k_e =0$ is incompatible with a periodic $\theta$.  We see that  in fact $k_e >0$ (with  our choice of conventions) since
$2\pi k_e = 2 \int d \vp e^{-\Phi}$.

We can express the action in terms of a diffeomorphism of the circle $f(t,\vp)$ with $f(t,\vp + 2\pi) = f(t,\vp) + 2\pi$ using the constraints and applying the field redefinition
\begin{equation}
\theta(t,\vp) = \frac{k_e(t)}{2} (f(t,\vp) - \vp)\,.
\end{equation}
The function $\cL$ now becomes
\begin{equation}
\cL = - \frac{k_e(t)}{4} f'{}^2 - \frac12 \{f, \vp \}\,,
\end{equation}
and the action is up to total derivative terms:
\begin{equation}\label{SgeomElliptic}
S_{\rm bdy}^{\Sigma_o} =  \frac{k}{8\pi} \int_{\Sigma_o} dt d\vp \; \left[ \frac{\partial_- f' f''}{f'{}^2} - k_e(t)^2 \partial_- f f' \right]\,.
\end{equation}
The result is once again phrased in terms of the geometric action of \cite{Alekseev:1988ce}, with holonomy enhanced to be dynamical and fulfilling $\dot{k}_e = 0$ on shell, but now 
\begin{equation}
b_0 = - \frac{c}{48\pi} k_e(t)^2 \,.
\end{equation}
The difference between this action and \eqref{Sgeom} is the relative sign of the representative term. So we have established that elliptic holonomies lead to negative orbit representatives. From here we could obtain the hyperbolic holonomy by analytic continuation $k_e = i k_0$. This implies that the action \eqref{SgeomElliptic}, when expressed as the chiral boson, would have purely imaginary zero modes.
The relevant Virasoro orbits are now $\textrm{Diff}(S^1)/S^1$, except for the particular values of the holonomy discussed in the next subsection.

\subsubsection{Gauge invariance}

The gauge invariance of the action \eqref{WZW2bdy} is parametrized by a time-dependent  $SL(2, \mathbb{R})$ element $\omega(t)$.  Just as in the hyperbolic case, this gauge invariance is partly fixed by the choice of the form of the holonomy and generically reduces to a $U(1)$ gauge symmetry.  In the elliptic case, a new feature arises: there are exceptional values for which this is not the case.  These correspond to $k_e = $ integer, in which case $K$ is equal to $K= \pm I$ and is an element of the center of $SL(2, \mathbb{R})$ commuting with all $\omega(t)$'s.  In that case, the direction of the axis of rotation is irrelevant and one can get rid of that information contained though the parametrization $K=\frac{k_e}{2}   (L_--L_+)$.

To see how this comes about, let us we take the $SL(2,\bR)$ element $\omega(t)$ to be
\begin{equation}
\omega(t) = \left( \begin{array}{cc}  d(t) & c(t) \\ b(t) & a(t) \end{array}\right)\, , \qquad a(t)d(t) - b(t)c(t) = 1\, ,
\end{equation}
then the field $\theta(t,\vp)$ in the Iwasawa decomposition transforms as
\begin{equation}
\tan \theta \to \tan \hat \theta(t,\vp)= \frac{a(t) \tan \theta(t,\vp) + b(t)}{c(t) \tan \theta(t,\vp)+ d(t)} \,.
\end{equation}
In terms of the field $f(t,\vp)$ the $SL(2,\mathbb{R})$ invariance becomes manifest when $k_e(t) = n$ and $n \in \mathbb{Z}$. In that case there is a transformation to a periodic field $\Theta$ defined as
\begin{equation}\label{trafo}
\Theta(t,\vp) = \tan \left(\frac{n}{2}f(t,\vp) \right) \,.
\end{equation}
In terms of $\Theta$ the action becomes 
\begin{equation}\label{Sgeomnohol}
S(\Theta) = \frac{k}{4\pi} \int dt d\vp \left[ \frac{\partial_t \Theta''}{\Theta'} - \frac32 \frac{\partial_t \Theta' \Theta''}{\Theta'{}^2}\right] \,.
\end{equation}
The transformation \eqref{trafo} has the effect of removing the representative term on the orbit, i.e., $k_e$. When trying to remove the orbit representative term for $k_e \neq n$, the same trick does not work as $\Theta$ is not periodic anymore, and the periodicity of $\Theta$ will reflect the value of the holonomy (orbit representative).

The resulting action (\ref{Sgeomnohol}) corresponding to $k_e = n$ is manifestly invariant under $SL(2,\bR)$. Indeed, taking in the action
\begin{equation}
\Theta \to \hat \Theta = \frac{a(t) \Theta(t,\vp) + b(t)}{c(t) \Theta(t,\vp)+ d(t)}\,,
\end{equation}
leads to 
\begin{align}
\hat S (\hat \Theta) = S(\Theta) - \frac{k}{4\pi} \int dt d\vp \; \Bigg\{ \partial_\vp \left( \frac{\partial_t \hat \Theta'}{\Theta '} \right) - \partial_t \left( \log (c \Theta + d) \partial_\vp \log \Theta' \right) \\
+ \partial_\vp \left( \log (c \Theta + d) \partial_t \log \Theta' \right) - 2 \partial_t \partial_\vp \log(c\Theta + d) + 2 \partial_\vp\left( \frac{\dot c}{c} \log (c \Theta +d) \right) \Bigg\} \nonumber \,.
\end{align}
The difference between the transformed action and the original one are only total derivative terms. The Virasoro orbits are thus  $\textrm{Diff}(S^1)/SL(2, \mathbb{R})$ in the degenerate case. 

When $k_e \not=n$, the gauge invariance is partially fixed  to $U(1)$ through the choice of the form of the holonomy.  The enhancement of the gauge symmetry for $k_e=n$ also manifests itself in the dimension of the space of solutions of the constraint equations, which is increased (see appendix \ref{app:elliptic}).

\subsection{Parabolic holonomy}\label{sec:parabolic}
The final case we should consider are holonomies in the parabolic conjugacy class of $SL(2,\mathbb{R})$. We will parameterize these by taking
\begin{equation}
K(t) = k_p(t) L_+ \,,
\end{equation}
observing that $k_p$ can be absorbed through redefinitions and should thus disappear from the final formulas as a consistency check.

To study the parabolic holonomy, we adopt the  Iwasawa decomposition \eqref{Iwasawa} but invert the order and write:
\begin{equation}
h = n_+.a.k\,.
\end{equation}
and likewise for $l$, with the appropriate field redefinition of the fields at inner boundary, such that $l = n_-.a.k$.

The action \eqref{WZW2bdy} again splits into two boundary actions:
\begin{equation}
S = S^{\Sigma_o}_{\rm bdy}[\theta,\Phi,\eta,k_p] - S^{\Sigma_i}_{\rm bdy}[\vartheta,\Psi,\nu,k_p] \,,
\end{equation}
with:
\begin{equation}
S_{\rm bdy}^{\Sigma_o}[\theta, \Phi,\eta,k_p] =  \frac{k}{4\pi} \int dt d\vp \, \left[ \frac12 \Phi' \partial_- \Phi - 2 \theta' \partial_- \theta + 2 e^{-\Phi} \partial_- \theta (\eta' + k_p(t)) \right]\,.
\end{equation}
The constraints from the boundary conditions now imply the following differential relations
\begin{equation}\label{paraconstraints}
\Phi' = 2 (\theta' - 1) \cot \theta \,, \qquad \eta' = - k_p - e^{\Phi} \Phi' \cot 2\theta \,,
\end{equation}
and 
\begin{equation}
\cL = - \theta' - \frac12 \Phi' \cot \theta
\end{equation}
Using the second relation in \eqref{paraconstraints} we see that the dependence on $k_p(t)$ drops out of the action
\begin{equation}\label{para}
S_{\rm bdy}^{\Sigma_o} =  \frac{k}{4\pi} \int dt d\vp \, \left[ \frac12 \Phi' ( \partial_- \Phi - 4 \partial_- \theta \cot 2\theta) - 2 \theta' \partial_- \theta  \right]\,.
\end{equation}
In order to express the action in terms of a single field, we must somehow find a way to integrate the first relation in \eqref{paraconstraints}. A useful way to do so is to change variables to a function $\phi(t,\vp)$ defined as
\begin{equation}\label{phidef}
\cot \theta = - \frac12 \phi' \,.
\end{equation}
Then we can integrate the first of the relations \eqref{paraconstraints} to obtain
\begin{equation}
\Phi = \phi - \log(4+ \phi'^2) \,.
\end{equation}
where we have dropped the integration constant $c(t)$, which reflects the gauge redundancy. This parameterization of $\theta$ is useful, because it brings the stress-tensor $\cL$ to a familiar form:
\begin{equation}
\cL = \frac14 \left( \phi'^2 - 2 \phi'' \right)\,.
\end{equation}
This is the form of the stress tensor for a chiral boson $\phi$. Indeed also the action \eqref{para} turns into a familiar form. Up to total derivatives, we find
\begin{equation}\label{parafin}
S_{\rm bdy}^{\Sigma_o} =  \frac{k}{4\pi} \int dt d\vp \, \left[ \frac12 \phi'  \partial_- \phi - \partial_- \phi' \right]\,.
\end{equation}
This is the chiral boson action, but now without holonomy contribution. Similar arguments hold on the other boundary.
From the chiral boson, the map to the geometric action \eqref{Sgeom} is performed by yet another field redefinition
\begin{equation}\label{fdef}
e^{\phi} = f'\,,
\end{equation}
such that the action becomes
\begin{equation}\label{Sgeomparabolic}
S_{\rm bdy}^{\Sigma_o} =  \frac{k}{8\pi} \int dt d\vp \, \frac{f'' \partial_- f'}{f'{}^2} \,,
\end{equation}
and likewise on the other boundary. This is the geometric action on the coadjoint orbit of the Virasoro group with vanishing representative. As shown in the previous section, in principle this action is invariant under the full $SL(2, \mathbb{R})$, however not all $f(t,\vp) \to \frac{a(t) f(t,\vp) + b(t)}{c(t) f(t,\vp) + d(t)}$ are compatible with the periodicity condition on $f(t,\vp)$ inherited by the field redefinitions \eqref{phidef} and \eqref{fdef}. This reduces the full $SL(2,\mathbb{R})$ to a one-dimensional abelian subgroup, consistent with the fact that we have partly fixed the gauge by choosing the holonomy to lie in the parabolic conjugacy class of $SL(2,\mathbb{R})$. 

The orbits are thus $\textrm{Diff}(S^1)/\mathbb{R}^1$ in the parabolic case.  Note that we do not find in our analysis the exceptional orbits that do not contain constant Virasoro charges because of our assumption (\ref{explicitAphi}), see footnote \ref{footnote3}.


\section{Conclusions}\label{sec:Conclusions}

The main result of this paper is the explicit derivation of the
gravitational action for three-dimensional gravity in the case when the
topology of the spatial sections is that of the annulus, with boundary
conditions expressing asymptotic $AdS_3$ behaviour at both boundaries.
The action \eqref{Shyper} is not a sum of boundary actions but involves in addition
couplings to the zero modes, which are additional degrees of freedom to
be taken into account.  These are the holonomy and one radial Wilson
line \eqref{radialWilson} connecting the boundaries (the other radial Wilson lines being
expressible in terms of any one of them and the boundary fields).  These
global degrees can (and must!) be consistently varied in the action
principle.

We also made the connection with the geometrical actions \eqref{Sgeom}, \eqref{SgeomElliptic} and \eqref{Sgeomparabolic}, and showed how the exceptional orbits with enhanced stability subgroups correspond to an enhanced gauge symmetry of the action and a degeneracy of the solution of the Drinfeld-Sokolov reduction constraints.

While we considered the specific example of two asymptotically anti-de
Sitter regions, most of our considerations  on holonomies and radial
Wilson lines qualitatively apply whenever there are two boundaries,
independently of the form that the boundary conditions explicitly take
there.  So, in particular, one boundary can describe an asymptotically
anti-de Sitter region and the other can describe a black hole horizon,
along the lines of \cite{Afshar:2016kjj,Grumiller:2019tyl} (see also \cite{Ojeda:2019xih,Grumiller:2019fmp}).  Although we
have not explored the problem, similar features are also expected to
hold with more boundaries.

The results of this paper apply equally well to supergravity models, because these are also Chern-Simons theories with boundary conditions of the Drinfeld-Sokolov type implementing a Hamiltonian reduction at the boundary \cite{Henneaux:1999ib}.  The resulting asymptotic symmetry algebras are the $\cN$-extended superconformal algebras of \cite{Knizhnik:1986wc,Bershadsky:1986ms,Fradkin:1992bz,Fradkin:1992km}, which are linear for $\cN \leq 2$.

In the annulus case, one can include explicitly the holonomies along the above lines.  This is most conveniently done by treating separately the two chiralities, leading, for each chirality,  to a supersymmetric chiral action at each boundary, coupled by radial Wilson lines.
One also finds that the system is physically described by two sets of generators of the superconformal algebras, one at each boundary.  These generators are constrained by the holonomy matching condition and provide, together with the global modes, a complete  description of the system.  The dynamics reduce to that of the dynamics of these generators and of the global modes, and can therefore be expressed in terms of geometrical actions. The details will be presented elsewhere \cite{InPreparation}.

There is one new ingredient that comes in, however, when the boundary algebras are nonlinear ($\cN >2$).  It is that the geometrical actions should not be formulated in terms of orbits of the coadjoint representation, since the phase space does not provide a linear representation, but rather in terms of the more general concept of symplectic leaves \cite{Lichne,Weinstein}.  The generators of the asymptotic symmetry algebra form a Poisson manifold, with a Poisson bracket  that is degenerate if one focuses only on a single boundary algebra without including the global radial Wilson lines.  The symplectic leaves of this Poisson manifold have a well-defined symplectic structure, which is the one that enters in the action.  
A similar phenomenon appears when including higher spins gauge fields, the asymptotic symmetry algebras of which are the non-linear $\cW$-algebras \cite{Henneaux:2010xg,Campoleoni:2010zq}.

It is well known that combining two chiral bosons of opposite
chiralities yields the Liouville theory (this is e.g.  recalled in the
appendix of \cite{Henneaux:1999ib} where references are given). This was
used in \cite{Coussaert:1995zp} to formulate the boundary theory as a
Liouville model, but the zero modes were not handled there. This
Liouville reformulation of $AdS_3$ gravity was the starting point of
\cite{Li:2019mwb}.  It would be interesting to investigate whether the
inclusion of holonomies and radial Wilson lines among the dynamical
variables would lead to more solutions and states than found in
\cite{Li:2019mwb}.

\vspace{.2cm}

\noindent
{\bf Note added:} While this work was completed,  the interesting
preprint \cite{Grumiller:2019ygj} came out, which also explicitly considers
two boundaries in $AdS_3$ gravity, but with different goals and
along different lines.

\section*{Acknowledgements}

 Work  partially supported by the ERC Advanced Grant ``High-Spin-Grav'' and by FNRS-Belgium (convention IISN 4.4503.15).

\vspace{.5cm}

\appendix

\begin{center}
\bf{\Large{Appendices}}
\end{center}

\section{Conventions}\label{app:conventions}
The $\mathfrak{sl}(2,\mathbb{R})$ generators are noted by $L_0$, $L_\pm$ and they satisfy the algebra
\begin{equation}\label{bracket-init}
[L_0,L_{\pm}] = \pm L_{\pm}\,, \qquad [L_+,L_{-}]=2L_0 \,.
\end{equation}
Thorough the paper we use the following matrix representation of $\mathfrak{sl}(2,\mathbb{R})$
\be
L_0 = \frac{1}{2}\left(\begin{array}{cc}
1 & 0 \\
0 & -1
\end{array}\right),\qquad L_+ = \left(\begin{array}{cc}
0 & 1 \\
0 & 0
\end{array}\right), \qquad L_- = \left(\begin{array}{cc}
0 & 0 \\
1 & 0
\end{array}\right).
\ee
They enjoy the property that
\be
\tr(L_0 L_0 ) = \frac{1}{2}\,,\qquad \tr(L_+ L_-)= \tr(L_- L_+) = 1 \,.
\ee

\section{From the infinite cylinder to the annulus}
\label{app:cylinder}

In this appendix, we consider the eternal black hole that has two asymptotic regions at infinity, so that one can formally say that ``$r^{\rm outer}$ is at $r\to \infty$ and $r^{\rm inner} $ is at $r \to - \infty$''.  We show how one can assume the ``radial'' coordinate to have a finite range.

To reproduce metrics with local AdS$_3$ asymptotics in the outer asymptotic region, we must asymptotically take  in that region \cite{Coussaert:1995zp}	
\begin{equation}
A_r = b(r)^{-1} \partial_r b(r), \quad A_\vp = b(r)^{-1} \left(L_- + \cL(t,\vp) L_+\right)b(r), \quad b(r) = \exp( L_0 \ln r)\,,
\end{equation}
in terms of the Schwarzschild radial coordinate $r$ of the outer region. The asymptotic $r$ dependence of the connection in the outer asymptotic region can therefore be thought of as being induced from the asymptotically constant connection with purely angular component $a_\vp = L_- + \cL L_+$ by an $r$-dependent gauge transformation.  This amounts to taking the group element that appears in the solution of the zero curvature condition to be of the form $G = g(t, \vp)) b(r)$.  If we undo that gauge transformation, the connection is simply $a_\vp$, and since it has no asymptotic $r$-dependence, we can trivially rescale the $r$-coordinate so that the new radial coordinate takes a finite value at the ``outer asymptotic boundary''.  We assume in the text that all these transformations have been performed and keep denoting the resulting connection $A_i$. 

Similarly, the connection in the inner asymptotic region takes the asymptotic form
\begin{equation}
\tilde{A}_\vp = \tilde{b}(\tilde{r})^{-1} \left(L_+ + \tilde{\cL}(t,\vp) L_-\right)\tilde{b}(\tilde{r})\,, \qquad \tilde{b}(\tilde{r}) = \exp( -L_0 \ln \tilde{r})\,,
\end{equation}
where $\tilde{r}$ is now the Schwarzschild radial coordinate of the inner region. One can thus repeat the same considerations as for the outer boundary. 

By making a gauge transformation by a group element that interpolates between $b(r)$ in the outer region and $\tilde{b}(\tilde{r})$ in the inner region and following the above procedure, one can simultaneously bring both boundaries at finite value of the radial coordinate and assume that the field takes at the boundaries the simple form given in the text.

\section{Consistency of the Hamiltonian reduction} 
\label{app:Consistency}

We consider in this appendix some aspects of the Hamiltonian reduction procedure that leads to the chiral boson action starting from the $SL(2, \mathbb{R})$ WZW model. 

 The equations obtained by varying \eqref{Sbdy} with respect to $\Phi$, $X$ and $Y$ are
\begin{subequations}
	\label{fulleom}
	\begin{align}
	\delta \Phi : && \partial_- \Phi' + \dot k_0 - 2 e^{\Phi} \partial_- X (Y' - k_0 Y) & = 0 \,, \\
	\delta X : && -\partial_-( 2 e^\Phi (Y' - k_0 Y)) & = 0 \,, \\
	\delta Y : && -\partial_\varphi( 2 e^\Phi \partial_- X )- 2 e^\Phi \partial_- X k_0 & = 0 \,.
	\end{align}
\end{subequations}
Of course, we have similar equations on the inner boundary, and one should also vary the action with respect to the dynamical holonomy $k_0$, which yields the time evolution of the conjugate to the holonomy,
\be
2 \pi (\dot{\Phi}_0 - \dot{\Psi}_0) = \oint d\vp \left(   2 e^{\Phi} Y \partial_- X - 2 e^{\Psi} V \partial_+ U \right) + 2k_0 \,.
\ee

By using the constraints \eqref{Gaussconstraints} in the field equations we see that the first two equations vanish trivially, while the last equation becomes (for $e^\Phi \neq 0$):
\begin{equation}\label{eom1}
\partial_- \left(X' + X^2\right) = 0\,.
\end{equation}
This equation is actually the statement that the stress tensor is holomorphic since it is equivalent to $\partial_- \cL = 0$.

By varying the action \eqref{Sred} with respect to $\Phi$ we obtain
\begin{equation}\label{eom2}
\partial_- \Phi' + \dot k_0 = 2 \partial_- X = 0\,.
\end{equation}
While it is clear \eqref{eom2} certainly implies \eqref{eom1}, the converse might at first sight seem not necessarily to be true.   However, it should be recalled that we are dealing with a  field $X$ defined on the circle, i.e., periodic. This makes  \eqref{eom2} and \eqref{eom1} equivalent.

To see this, we observe that the equation \eqref{eom1} implies that
\be
X' +X^2 = b(x^+)\,,
\ee
where $b(x^+)$ is an arbitrary function of $t + \varphi$ but does not depend on $x^- = t - \vp$.

This equation is a first order differential equation for $X$.  According to general theorems, the solution is unique on the real line up to an integration constant, which can be taken to be the value of $X$ at $\varphi = 0$, or which can be parametrized in any other convenient way.  On the circle, the solution is also unique, but this time {\em at most} up to an integration constant, because the integration constant must be compatible with the fact that $X$ should be periodic, a requirement that might fix it and make the solution unique.

Our goal is to show that this is actually the case.  Instead of considering the general case, which was actually implicitly treated in \cite{Henneaux:1999ib} where it was shown that it was legitimate to insert the Hamiltonian reduction constraints inside the action, we consider in turn two illustrative cases, to exhibit the mechanism that makes the equations equivalent:
\begin{itemize}
	\item $b =$ constant = $c^2$ ($b >0$ because $2 \pi b = \int d \varphi x^2 \geq 0$  and  if $b = 0$, then clearly $x=0$ and involves no integration constant)
	\item $b = c^2 + \epsilon(\varphi)$, where the periodic function $\epsilon$ is small with respect to $b$.
\end{itemize}

So, consider first the case $b = c^2$ with $c>0$. Since $X(\varphi)$ is periodic, its derivative vanishes somewhere, say at $\varphi_0$.  One has $X(\varphi_0) = c$ or $X(\varphi_0) = - c$. Consider the differential equation $X' + X^2 = c^2$ with initial condition $X = c$ (or $-c$) at $\varphi_0$.  Its solution is unique (it is first order and its initial condition $X(\varphi_0) = c$ or $-c$ is given).  This unique solution is easily verified to be $X = c$ (or $-c$).  The reference to $\varphi_0$ accordingly completely disappears, i.e., one gets the same solution no matter what $\varphi_0$ is.  Therefore, $X$ is uniquely determined by $b$ in this case, up to a sign.  This implies $\partial_- X = 0$ since $X(x^-)$ must be equal to $\pm c$ and cannot jump from one value to the other during the time evolution (by continuity).

Turn now to the case $b = c^2 + \epsilon(\varphi)$ and write $X = c + \eta(\varphi)$.  The equation becomes, neglecting squares of small quantities:
\be
\eta' + 2 c \eta = \epsilon \,.
\ee
The general solution can be obtained by the method of variations of constants and read:
\be
\eta(\varphi) = \left( \int_0^\varphi d \theta \epsilon(\theta) e^{2c\theta} + K \right) e^{-2c \varphi}.
\ee
Periodicity fixes the integration constant $K$ to be
\be
K = \frac{\int_0^{2 \pi} d \theta \epsilon(\theta) e^{2c\theta} }{1 - e^{-4 \pi c}}
\ee
(the denominator does not vanish since $c \not=0$).  The solution $X$ is therefore unique and since $\epsilon$ does not depend on $x^-$, one gets again $\partial_- X = 0$.

\section{Elliptic holonomy and enhancement of the constraint solutions}
\label{app:elliptic}

We wish to investigate here how unique (up to a gauge transformation) the solution $(\Phi(\vp), \theta(\vp), \eta(\vp))$ of the constraints \eqref{constraintell} for $\cL(\vp)$ given by \eqref{Lell} is.

To gain understanding on this problem, let us assume to begin with that $\cL$ is a constant, $\cL = L$.
We first try to determine the constant solutions to the problem. We denote by $P$ the
constant value of $\Phi$. The field $\eta$ is necessarily zero, $\eta = 0$, while the constant value of
$\theta$ can be shifted by a gauge transformation, so we can assume $\theta = 0$.
One clearly has
\begin{equation}
L = - e^{-2P}\,, \qquad k_e = 2 e^{-P} = 2 \sqrt{-L}\,,
\end{equation}
so in particular
\begin{equation}
L = - \frac14 k_e^2 \,.
\end{equation}
Anti-de Sitter space corresponds to $k_e = 1$ and $L = - \frac14$.

Are there other solutions (with same given $L$ and hence $k_e$)? To explore this
question, we perturb around the solution just derived,
\be
\Phi(\vp) = P + p(\vp)\,.
\ee
The fields $\theta$ and $\eta$, having zero background value, coincide with the perturbations. The
perturbed equations read
\be
\theta' + p \frac{k_e}{2} = 0\,,
\qquad \eta = p' \,, \qquad
0 = k_e^2 p + p'' \,.
\ee
The general solution for $p$ is
\be
p = \alpha \cos(k_e \vp) + \beta \sin(k_e \vp)\,,
\ee
but it will not be periodic when $k_e$ is not an integer, unless one takes $\alpha = \beta = 0$.
Hence, in that case, $p = 0, \eta = 0, \theta = $constant (can be absorbed by a residual gauge
symmetry). The solution is unique.

When $k_e$ is an integer, however, the solution $p$ is periodic for any choice of $\alpha$ and $\beta$.
So there are two more families of solutions in addition to the constant one. (Given $p, \theta$ exists and
is unique up to a gauge transformation because $\theta' = - \frac{k_e}{2} p = \frac{1}{2k_e} p''$ and so $\theta = \frac{1}{2k_e} p' + $
constant.)

There is thus an enhancement of the number of independent solutions when $k_e$ is an integer. This is precisely the values of $k_e$ for
which there is enhanced gauge symmetry. The nonlinear treatment going beyond the perturbative treatment involves the Hill
equation.



\providecommand{\href}[2]{#2}\begingroup\raggedright\endgroup

\end{document}